\documentclass[lettersize,journal]{IEEEtran}
\usepackage{amsmath,amsfonts}
\usepackage{algorithmic}
\usepackage{algorithm}
\usepackage{array}
\usepackage{textcomp}
\usepackage{stfloats}
\usepackage{url}
\usepackage{verbatim}
\usepackage{graphicx}
\usepackage{subcaption}
\usepackage{makecell}
\usepackage{hyperref}
\usepackage{cite}
\usepackage{multirow}
\usepackage{orcidlink}
\usepackage[T1]{fontenc}
\usepackage{tcolorbox}
\definecolor{boxblue}{RGB}{230, 242, 252}
\usepackage{changepage}
\usepackage{framed}
\makeatletter
{\par\unskip\endMakeFramed}
\makeatother
\definecolor{formalshade}{rgb}{0.93,0.93,0.93}
\definecolor{darkblue}{rgb}{0.2, 0.2, 0.2}
\newenvironment{formal}{%
\def\FrameCommand{%
  \hspace{1pt}%
  {\color{darkblue}\vrule width 2pt}%
  {\color{boxblue}\vrule width 4pt}%
  \colorbox{boxblue}%
}%
\MakeFramed{\advance\hsize-\width\FrameRestore}%
\noindent\hspace{-1pt}
\begin{adjustwidth}{}{7pt}%
\vspace{2pt}\vspace{2pt}%
}
{%
\vspace{3pt}\end{adjustwidth}\endMakeFramed%
}
\newcounter{resultcounter}

\newcounter{patterncounter}

\def\BibTeX{{\rm B\kern-.05em{\sc i\kern-.025em b}\kern-.08em
    T\kern-.1667em\lower.7ex\hbox{E}\kern-.125emX}}
    
\usepackage{tcolorbox}
\tcbuselibrary{skins, breakable}
\newtcolorbox{summarybox}[2][]{
  colback=gray!10,
  colframe=black!75,
  fonttitle=\bfseries,
  title=#2,
  breakable,
  enhanced jigsaw,  
  pad at break*=2mm, 
  #1
}

\begin{document}

\title{Automating the Detection of Requirement Dependencies Using Large Language Models}

\author{\IEEEauthorblockN{Ikram Darif\IEEEauthorrefmark{1}\orcidlink{0009-0002-2646-4471},
Feifei Niu\IEEEauthorrefmark{1}\orcidlink{0000-0002-4123-4554}, Manel Abdellatif\IEEEauthorrefmark{2}\orcidlink{0000-0002-8647-1676}, Lionel C. Briand\IEEEauthorrefmark{1}\IEEEauthorrefmark{3}\orcidlink{0000-0002-1393-1010}, 
Ramesh S\IEEEauthorrefmark{4}\orcidlink{0000-0002-8501-7447}
and
Arun Adiththan\IEEEauthorrefmark{4}\orcidlink{0009-0004-4586-8417}}
\IEEEauthorblockA{\\
\IEEEauthorrefmark{1} University of Ottawa, Ottawa, Canada. \\
\IEEEauthorrefmark{2} École de technologie supérieure, Montreal, Canada. \\
\IEEEauthorrefmark{3} Research Ireland Lero Centre for Software Research, University of Limerick, Ireland.\\
\IEEEauthorrefmark{4} General Motors, Detroit, Michigan, United States. \\}
Email: idarif@uottawa.ca; fniu2@uottawa.ca; manel.abdellatif@etsmtl.ca; lbriand@uottawa.ca; \\
ramesh.s@gm.com; arun.adiththan@gm.com.
}





\maketitle

\begin{abstract}
Requirements are inherently interconnected through various types of dependencies. Identifying these dependencies is essential to managing change, as they underpin critical decisions and influence a range of activities throughout the software development life cycle. However, this task is challenging, particularly in modern software systems, given the high volume of complex, coupled requirements. These challenges are further exacerbated by the inherent ambiguity of Natural Language (NL) requirements and their constant change. Consequently, requirement dependency detection is often overlooked or performed manually. Large Language Models (LLMs) show strong NL processing capabilities, presenting a promising avenue for requirement-related tasks. While they have been shown to enhance various requirements engineering activities, their effectiveness at identifying requirement dependencies remains underexplored. 
In this paper, we introduce LEREDD, an LLM-based approach for automated requirement dependency detection that leverages Retrieval-Augmented Generation (RAG) and In-Context Learning (ICL). It identifies diverse dependency types directly from NL requirements. We compare various state-of-the-art (SOTA) LLMs and prompting strategies for requirement dependency detection. Furthermore, 
we empirically evaluate LEREDD across four datasets, including one from industry, comprising a total of 929 requirement pairs. We compare LEREDD against two SOTA baselines, considering both intra-system and inter-system few-shot example selection. The results indicate that LEREDD outperforms the baselines overall, particularly in detecting specific dependency types, yielding average relative gains of 124.4\% and 16.9\% in $F_1$ scores over the two baselines, respectively, for the intra-system setting, and 66.67\% and 280\% for the inter-system setting. Furthermore, LEREDD achieves highly accurate classification for non-dependent requirements, with an average $ F_1$ score of 93\% across all experiments. This is particularly significant because non-dependent pairs typically constitute the vast majority in real-world specification documents. By reliably filtering these cases, LEREDD substantially reduces the manual effort and time required for dependency analysis. 
Industrial experts also exhibited a high level of agreement with LEREDD's predictions when evaluated on an industrial dataset, further supporting its applicability in industrial contexts and the reliability of its predictions.
Finally, we contribute an annotated dataset of requirement dependencies to support future research.

\end{abstract}


\section{Introduction}

Modern software systems are becoming increasingly complex, driven by growing and more advanced stakeholder and user demands. This complexity is directly reflected in Requirements Engineering (RE) processes, where requirements are systematically identified, documented, analyzed, and managed to accurately capture stakeholders’ needs~\cite{8559686}. As the foundational artifacts of the software development life cycle, requirements are arguably the most critical artifacts for project success and the quality of the final product~\cite{936219}. 

However, modern software systems are characterized by large numbers of requirements, high requirement complexity, and strong interdependencies~\cite {10.1007/978-3-540-69062-7_11}. These requirements are often developed and maintained over long periods, involving hundreds of engineers, particularly in regulated domains subject to functional safety standards. Indeed, requirements are not independent entities. They are inherently interconnected through various kinds of dependencies~\cite {Motger2025}. Identifying, classifying, and managing inter-requirement dependencies is crucial, as they underpin critical software development decisions and influence activities throughout the software development life cycle~\cite{Motger2025}. For instance, they are essential for conducting accurate impact analysis during system changes, ensuring consistency by identifying conflicting requirements, and facilitating software development and testing. Ignoring such dependencies not only harms project success but also compromises release quality and leads to substantial rework~\cite{8920671,9218190}. However, detecting requirement dependencies remains error-prone, time-consuming, and cognitively challenging~\cite{Motger2025}. These challenges largely stem from the ambiguity introduced by the predominant use of Natural Language (NL) to author requirements, as well as from reliance on manual effort for detection.

Existing approaches for detecting requirement dependencies exhibit various limitations. Retrieval-based approaches~\cite{10759654,10089332,li-etal-2015-recovering,8995268,math12091272} are restricted to pairwise classification and rely on fixed representations, failing to account for domain-specific context. Knowledge-based approaches use ontologies or graphs to represent domain knowledge~\cite{8874920,9315489,Guo2021AutomaticallyDT,10.1007/978-3-030-73128-1_3,https://doi.org/10.1002/sys.21461,Motger2019OpenReqDDAR,9218190}, but require substantial effort for development and maintenance~\cite{Motger2025}. Finally, ML-based approaches~\cite{MohamedEtAl2025,Graler_Oleff_Hieb_Preuss_2022,10.1007/978-3-030-73128-1_2,Fischbach2021CiRAAT,8920671,8609673,101007,10.1007/978-3-030-93709-6_29} require large training datasets and struggle with the class imbalance inherent in dependency detection, as independent requirements typically far outnumber dependent ones~\cite{Motger2025}. 

Large Language Models (LLMs) have significantly revolutionized artificial intelligence and its applications, emerging as advanced models with billions of parameters trained on vast corpora. Their ability to be fine-tuned for specialized applications without exhaustive task-specific training has contributed to their success~\cite{MohamedEtAl2025,wu2024continuallearninglargelanguage}. They are particularly renowned for their capabilities in natural language processing, reasoning, and generation, rendering them especially relevant for natural language requirement engineering tasks~\cite{10.1145/3744746,zadenoori2025largelanguagemodelsllms}. While LLMs have been successfully applied to various RE-related tasks, notably requirements elicitation and classification, their utility for automated requirement dependency detection remains underexplored compared with other requirements engineering activities~\cite{fi16060180,zadenoori2025largelanguagemodelsllms}. Furthermore, the few existing LLM-based dependency detection approaches are either specific to certain application domains or limited to particular dependency types, limiting their applicability.

Leveraging LLMs' natural language processing capabilities, we propose LEREDD: LLM-Enabled REquirement Dependency Detection, an automated approach for detecting various types of direct dependencies between NL requirement pairs. LEREDD utilizes In-Context Learning (ICL) to dynamically retrieve relevant examples for each dependency type and the no-dependency case and leverages Retrieval-Augmented Generation (RAG) to extract domain-specific context from the Software Requirements Specification (SRS) document if the retrieved examples originate from a different system. The retrieved examples provide a comprehensive, domain-specific context that effectively guides the LLM during detection. LEREDD represents a novel contribution, as it is the first approach to investigate Retrieval-Augmented Generation (RAG) and few-shot learning, specifically semantic similarity-based dynamic example retrieval, for requirement dependency detection. It is also the first LLM-based approach to support the detection of multiple dependency types without being tailored to a specific application domain. For each requirement pair, LEREDD generates: (1) a prediction specifying the dependency type (e.g., \textit{Requires} and \textit{Implements}), (2) the rationale behind the prediction, and (3) a confidence score.
We investigate four LLMs, both proprietary and open-source, for detecting requirement dependencies. 

We empirically evaluate LEREDD on four systems, one of which is an industrial system, covering a total of 929 manually annotated requirement pairs. Specifically, we investigate six research questions:

\textit{\textbf{RQ.1: Which SOTA LLM is most effective for requirement dependency detection?}} We evaluate multiple SOTA LLMs under a zero-shot prompting setting to identify the most effective model for dependency detection.

\textit{\textbf{RQ.2: What is the optimal configuration for requirement dependency detection?}} We investigate different configurations for different prompting strategies and identify the best-performing configuration for each strategy, which is subsequently adopted in LEREDD.

\textit{\textbf{RQ.3: What are the contributions of contextual retrieval and dynamic example retrieval to LEREDD's performance?}} We compare LEREDD's performance with and without these stages, as well as with alternative few-shot example selection strategies.

\textit{\textbf{RQ.4: How does LEREDD compare with baseline approaches for intra-dataset dependency detection?}} We compare LEREDD with two baseline approaches in terms of detecting both \textit{No dependency} and specific dependency types when the few-shot examples originate from the same system.

\textit{\textbf{RQ.5: How does LEREDD compare with baseline approaches for cross-dataset dependency detection?}} We compare LEREDD with two baseline approaches in detecting both \textit{No dependency} and specific dependency types when the few-shot examples are drawn from a different system.

\textbf{\textit{RQ.6: To what extent do industrial experts agree with the predictions of LEREDD?}} 
We measure the degree to which industrial experts agree with the dependency predictions of LEREDD on an industrial dataset. 



The results show that using GPT-5.4 with zero-shot prompting outperforms open-source models for requirement dependency detection, reaching an average $F_1$-score of 78\%. However, while it achieves a high $F_1$-score of 87\% for detecting \textit{No dependency}, it struggles with specific dependency types, such as \textit{Requires} and \textit{Implements}, achieving an overall $F_1$-score of only 29\%, highlighting the limitations of zero-shot dependency detection. Evaluating different few-shot prompting and RAG configurations identified SBERT with Euclidean distance, average similarity aggregation, six examples per dependency type, and threshold-based re-annotation as the optimal few-shot configuration, whereas the optimal RAG configuration uses six 1000-character chunks. We then adopted these configurations in LEREDD. The ablation study of LEREDD shows that dynamically retrieving semantically relevant examples makes the most substantial contribution to its dependency detection performance. In contrast, adding contextual information through RAG provides no additional performance benefit over few-shot prompting alone in the intra-dataset setting.

Compared with baselines, LEREDD achieves highly accurate classification of non-dependent requirements, with an average $F_1$-score of 93\%. This is particularly significant because non-dependent pairs typically make up the vast majority of real-world SRS documents. By reliably filtering these cases, LEREDD substantially reduces the manual effort and time required for dependency analysis.
LEREDD also consistently outperforms the baseline approaches, particularly in detecting specific dependency types. For example, it achieves average relative gains of 124.4\% and 16.9\% in $F_1$-score across specific dependency types compared with the baselines in the intra-dataset setting. In the inter-dataset setting, incorporating RAG further improves LEREDD's performance, particularly for specific dependency types. LEREDD demonstrates strong inter-dataset robustness, with its overall $F_1$-score decreasing by only 2.27\% compared with the intra-dataset setting. It achieves relative gains in overall $F_1$-scores of 22.86\% and 50.88\% over the baselines, with the gains for specific dependency types increasing to 66.67\% and 280\%, respectively.

Industrial experts showed strong agreement with LEREDD's predictions, both in identifying the existence of dependencies and in determining their specific types, demonstrating the reliability of its results on industrial requirements. Furthermore, the rationales and confidence scores generated by LEREDD facilitated understanding and analysis of its predictions, supporting verification of its results.

Overall, LEREDD achieves high accuracy and $F_1$-scores, averaging 87\% and 86\%, respectively, across all dependency classes and the evaluated systems.
Moreover, LEREDD effectively addresses the inherent class imbalance among dependency types, achieving higher accuracy and $F_1$-scores than the baselines, particularly for the less frequent dependency types. LEREDD also shows strong robustness in cross-system evaluations, maintaining superior $F_1$ performance even when examples are retrieved from different systems. This capability is particularly valuable in real-world settings, where annotated data from the target system may be unavailable. 
Furthermore, LEREDD produced predictions that were similar to those of industrial experts, demonstrating the correctness of its results on industrial requirements.
We release the annotated corpus used in this study as an open-source dataset for benchmarking and training, helping to mitigate the scarcity of public datasets.

The remainder of the paper is organized as follows. Section~\ref{sec:relatedWork} reviews related work on requirement dependency detection. Section~\ref{sec:problem} presents the problem statement and discusses the importance of the different dependency types. Section~\ref{sec:approach} outlines the LEREDD framework. Section~\ref{sec:empiricalEvaluation} and Section~\ref{sec:Results} describe the empirical setup and results of our evaluations, respectively. Section~\ref{sec:discussion} provides an in-depth discussion of the findings, while Section~\ref{sec:threatsToValidity} examines the threats to validity. Finally, Section~\ref{sec:Conclusion} concludes the paper and outlines future work.

\section{Related Work}\label{sec:relatedWork}

Several approaches have been proposed to detect dependencies between requirements. A common characteristic of such methods is their reliance on Natural Language Processing (NLP) techniques as an initial step for preprocessing and knowledge representation~\cite{Motger2025}. Since SRS are predominantly authored in Natural Language (NL), NLP supports the development of a structured knowledge representation based on the syntactic features and semantic knowledge from the SRS corpus~\cite{MohamedEtAl2025}. Based on their detection logic, requirement dependency detection approaches can be classified into four primary categories~\cite{Motger2025,MohamedEtAl2025}: information retrieval-based approaches~\cite{10759654,10089332,li-etal-2015-recovering,8995268,math12091272}, knowledge-based approaches~\cite{8874920,9315489,Guo2021AutomaticallyDT,10.1007/978-3-030-73128-1_3,https://doi.org/10.1002/sys.21461,Motger2019OpenReqDDAR,9218190}, ML-based approaches~\cite{MohamedEtAl2025,Graler_Oleff_Hieb_Preuss_2022,10.1007/978-3-030-73128-1_2,Fischbach2021CiRAAT,8920671,8609673,101007,10.1007/978-3-030-93709-6_29}, and LLM-based approaches~\cite{Gaertner2024ASE,app15189891,niu2026tvrautomotiverequirementtraceability}. 

\subsection{Information Retrieval-based Approaches}

Information retrieval-based approaches involve applying predefined rules to identify potential dependencies in a corpus~\cite{Motger2025}. Such approaches include linguistic- and vector-based methods~\cite{Motger2025}. Linguistically based methods leverage extended syntactic and semantic features, using techniques such as token-based cross-reference detection and pattern matching, to identify semantically related requirements~\cite{Motger2025}. For example, Sarwosri~\textit{et al.}~\cite{10759654} proposed an approach that combines clustering algorithms with a rule-based system to detect conflicts among functional requirements. OGAWA~\textit{et al.}~\cite{10089332} 
applied a pattern matching technique, where they used a set of domain-specific keywords to detect dependencies, followed by a sentence simplification phase to filter out false positives. 

Vector-based approaches rely on vector representations of requirements to build lexical- or semantic-based vector space models for the requirements corpus \cite{Motger2025}. Lexical approaches identify dependencies using statistical methods, with TF-IDF (Term Frequency-Inverse Document Frequency) commonly applied to assess term importance relative to the corpus~\cite{Motger2025}. Lexical methods are often paired with semantic methods, such as Latent Semantic Analysis, Universal Sentence Encoder, and Word2Vec. These methods generate requirement embeddings and apply statistical measures, such as similarity analysis, to identify dependencies. Because of their established efficacy, vector-based models are frequently used as evaluative baselines. 
For example, Li~\textit{et al.}~\cite{li-etal-2015-recovering} utilized TF-IDF paired with cosine similarity analysis as an unsupervised learning baseline for requirements traceability link detection. For binary requirement dependency detection, Samer~\textit{et al.}~\cite{8995268} combined TF-IDF and LSA, while Guan~\textit{et al.}~\cite{math12091272} implemented optimized TF-IDF variants. 

While vector-based approaches have traditionally been adopted for requirement dependency detection, they are limited to pairwise classification and rely on fixed representations that fail to account for domain-specific terminology. Furthermore, because requirements often contain inconsistent terminology and dependencies derive from the underlying system architecture, detection \textit{Requires} inferential reasoning that these approaches lack. 

\subsection{Knowledge-based Approaches}

Knowledge-based approaches utilize structured representations of domain knowledge to infer dependencies between requirements. Such approaches are generally classified into two categories: (1) graph-based methods, which leverage structural relationships within the requirement corpus, and (2) ontology-based methods, which utilize formal ontologies for domain representation. 

Several graph-based approaches have been proposed in the literature ~\cite{8874920,9315489,Guo2021AutomaticallyDT,10.1007/978-3-030-73128-1_3,https://doi.org/10.1002/sys.21461}. For example, Priyadi~\textit{et al.}~\cite{8874920} utilized text analysis to extract requirement dependencies, such as ``similar'', ``part-of'', and ``elaborate'' from SRS documents and model the requirement dependency graph. Asyrofi~\textit{et al.}~\cite{9315489} applied NLP techniques to model various dependency types, such as PART OF, AND, OR, and XOR. Beyond simple linking, Guo~\textit{et al.}~\cite{Guo2021AutomaticallyDT} applied finer semantic analysis to graph structures to automatically detect ``conflict'' dependencies. Similarly, Schlutter~\textit{et al.}~\cite{10.1007/978-3-030-73128-1_3} proposed an NLP pipeline that maps requirements into a semantic relation graph from which dependencies are identified. Mokammel~\textit{et al.}~\cite{https://doi.org/10.1002/sys.21461} also applied NLP to identify links, similarities, and contradictions between requirements, then clustered them using a graph representation. 

In requirement dependency detection, ontologies are often paired with ML ~\cite{Motger2019OpenReqDDAR, 9218190}. A prominent example is OpenReq-DD \footnote{\url{https://openreq.eu/tools-data/}}, a specialized tool that automatically detects requirement dependencies by using ontologies to capture domain-specific term relationships, combined with NLP and ML techniques~\cite{Motger2019OpenReqDDAR}. Building on this, Deshpande~\textit{et al.}~\cite{9218190} compared OpenReq-DD with an active learning approach for detecting ``requires'' and ``refines'' dependencies and proposed a hybrid framework that combines both approaches. While ontology-based approaches support more effective domain modeling than graph-based approaches, building and maintaining an ontology is labor-intensive, time-consuming, and heavily dependent on expert knowledge~\cite{Motger2025}.

\subsection{ML-based Approaches}

ML-based approaches use machine learning and deep learning models to predict potential dependencies in requirements, learning patterns from training data~\cite{Motger2025,MohamedEtAl2025}. 
For instance, Gräßler~\textit{et al.}~\cite{Graler_Oleff_Hieb_Preuss_2022} used fine-tuned BERT models to identify ``refines'' and ``requires'' dependencies. Similarly, Fischbach~\textit{et al.}~\cite{10.1007/978-3-030-73128-1_2,Fischbach2021CiRAAT} proposed the CiRA approach, utilizing fine-tuned BERT to detect causality between requirements. Deshpande~\textit{et al.}~\cite{8920671} relied on supervised learning to classify dependency types. They trained multiple classifiers (e.g., Naive Bayes and Random Forest) on an initial dataset and then used them to generate pseudo-annotations from unlabeled samples \cite{8920671}. To automate the identification of ``requires'' dependencies, Atas~\textit{et al.}~\cite{8609673} proposed an approach that leverages supervised classification techniques. They evaluated the approach using a variety of classifiers, with Random Forest achieving the highest performance. Guan~\textit{et al.}~\cite{101007} introduced an active learning algorithm that iteratively selects the most relevant requirement pairs for manual annotation, which are continuously used to refine the model~\cite{101007}. Abeba~\textit{et al.}~\cite{10.1007/978-3-030-93709-6_29} specifically targeted ``conflict'' dependencies in non-functional requirements, showing that a Bi-LSTM (bidirectional long short-term memory) model with pre-trained word2vec outperformed other classifiers.

Despite their potential, ML-based approaches are largely limited to pattern recognition and lack the inference reasoning needed for dependency detection. They rely heavily on large, high-quality annotated data, which is scarce in real-world contexts. Furthermore, traditional ML-based approaches often struggle with data imbalance, as independent requirements typically far outnumber dependent ones.

\subsection{LLM-based Approaches}


Current research on LLMs for requirement dependency detection is limited~\cite{fi16060180,zadenoori2025largelanguagemodelsllms}. G\"{a}rtner~\textit{et al.}~\cite{Gaertner2024ASE} introduced ALICE, a framework that combines LLMs with formal logic to identify requirement contradictions. Using a contradiction taxonomy and a decision tree, ALICE  outperformed LLM-only approaches~\cite{Gaertner2024ASE}. Similarly, Almoqren~\textit{et al.}~\cite{app15189891} integrated Knowledge Graphs with LLMs to identify requirement dependencies, such as \textit{requires} and \textit{conflicts}, from mobile app reviews. Using BERT in an LLM-driven active learning loop, they captured semantic and structural relations and achieved high precision~\cite{app15189891}. For trace link recovery, Niu~\textit{et al.}~\cite{niu2026tvrautomotiverequirementtraceability} introduced TVR, an approach that leverages RAG-enhanced LLMs to validate and recover traceability between stakeholder and system requirements. 

LLMs are particularly relevant for requirement dependency detection because they leverage inferential reasoning while enabling automation. Compared with ML approaches, LLMs reduce or eliminate annotation overhead and are less sensitive to data imbalance. Despite these advances, LLM-based requirement dependency detection remains scarce, with existing approaches mainly targeting specific application contexts (e.g., approaches based on formal logic~\cite{Gaertner2024ASE} or applied to mobile app reviews~\cite{app15189891}) and often confined to specific dependency types (e.g., contradictions~\cite{Gaertner2024ASE}). Recent literature explicitly advocates for further investigation of LLMs in this context~\cite{Motger2025}. Our research addresses this critical gap by proposing an LLM-based approach that automates this task while detecting a wide spectrum of dependency types.

\section{Problem Statement}\label{sec:problem}

Detecting requirement dependencies facilitates and influences various activities throughout the software development life cycle~\cite{Motger2025}. For instance, detecting requirement dependencies enables accurate change impact analysis by identifying all requirements affected by a proposed change, reducing the risk of overlooking necessary modifications. It also enables early detection of requirement inconsistencies and conflicts, reducing the time and effort required to correct errors later in the development process. Furthermore, dependency detection enhances traceability by establishing explicit links among related requirements and, in turn, other development artifacts. For safety-critical systems, it further supports standards compliance and certification by maintaining end-to-end traceability, thereby facilitating audits and demonstrating compliance with certification requirements. However, existing approaches still require manual effort. 
Although LLM-based requirement dependency detection shows promise for automating this task, research in this area remains limited, as discussed above, and existing approaches often focus on a single dependency type.

In this paper, we cover five dependency types: \textit{Requires}, \textit{Implements}, \textit{Conflicts}, \textit{Details}, and \textit{Is similar}. We derived these types from a SOTA classification~\cite{Motger2025}, except \textit{Implements}, which we introduced specifically to address our industrial partner's needs. We reused some definitions from a handbook~\cite{Motger2025} and refined others to reduce ambiguity and improve clarity for the LLM (e.g., \textit{Details} and \textit{Requires}). \autoref{tab:DefinitionsAndExamples} reports the definitions utilized in the prompt, along with illustrative examples for each dependency type. We extracted the examples from one of the datasets used in the evaluation. 

\begin{table*}[!htbp]
\fontsize{6pt}{7pt}\selectfont
\centering
\caption{Dependency Types Definitions and Illustrative Examples}
\label{tab:DefinitionsAndExamples}
\resizebox{\textwidth}{!}{
\begin{tabular}{p{2cm}|p{4cm}|p{5cm}}
\hline
 \textbf{Dependency Type} & \textbf{Definition} & \textbf{Example} \\ \hline

 \textit{\textbf{Requires}} & if the fulfillment of one requirement is a prerequisite to the fulfillment of the other requirement. & \textit{``If a car in an oncoming lane is within 500 feet of the ADB system, or a car in the same lane is within 250 feet, the subsystem shall decrease light intensity by 50\%.''} requires \textit{``The subsystem shall check for cars and road elements every 10 milliseconds and determine whether the subsystem shall decrease light intensity.''}\\ \hline

\textit{\textbf{Implements}} & if one is a higher-level requirement (e.g., a system or subsystem level requirement) that is fulfilled by the other lower-level requirement (e.g., a subsystem or component level requirement). & \textit{``The system shall verify communication between subsystems.''} implements \textit{``The subsystem shall have checks to mitigate the chances of a cybersecurity attack.''}\\ \hline

\textit{\textbf{Conflicts}} & if fulfilling one requirement restricts fulfillment of the other requirement. & \textit{``The system shall not irritate the driver with more than 3 updates or alerts in 10 minutes for the system.''} conflicts with \textit{``The system shall alert the driver if any malfunctions or errors have been detected within 5 seconds.''}\\ \hline

\textit{\textbf{Details}}& if both requirements describe the same action under the same condition, and one requirement provides additional details specifically regarding the shared action. & \textit{``The system shall respond to the surroundings around it in less than 2.5 seconds by dimming the light intensity 50\%.''} details \textit{``The system shall gradually change the light intensity to ensure a smooth operation for the driver by decreasing the intensity of the light by no greater than 50\% in one update.''} \\ \hline
 
\textit{\textbf{Is similar}}& if one requirement replicates partially or totally the content of the other requirement, resulting in redundancy. & \textit{``The system shall authenticate updates.''} is similar to \textit{``The system shall verify the source of the update.''}\\ \hline



\end{tabular}}
\end{table*}

Each dependency type is important and must be identified. The \textit{Requires} dependency is one of the most frequent dependency types, and its identification is essential because it determines the order in which requirements should be implemented. Specifically, the requirement that is required by another requirement should be implemented before the dependent requirement. Identifying such dependencies is also useful in multi-sprint software development, as it helps determine which requirements to implement in each sprint. Furthermore, identifying these dependencies is valuable for testing, as the required requirement can be treated as a precondition in test cases validating the dependent requirement. The \textit{Implements} dependency describes the implementation details associated with high-level requirements. Identifying this dependency supports software traceability, development, and design by clarifying how high-level requirements are decomposed into implementable low-level requirements. It also facilitates change impact analysis by identifying which requirements to review when a high-level requirement changes. Conversely, when a lower-level requirement changes, the corresponding high-level requirement should be reassessed to ensure that its intended functionality remains satisfied. 

The \textit{Conflicts} dependency identifies requirements that restrict or interfere with one another, making it valuable for early consistency checking. It also supports change impact analysis, as modifying one necessitates verifying the validity of the other. The \textit{Details} dependency links complementary requirements, supporting a more complete understanding of a functionality and facilitating its correct implementation. It can also reduce redundancy, as complementary requirements may be merged into a single, complete requirement. The \textit{Is Similar} dependency identifies semantically similar or redundant requirements, enabling the detection and removal of duplication within the specification. 

For safety-critical systems, these dependencies are particularly useful for supporting certification and compliance with applicable standards. For instance, \textit{Requires} dependencies help verify that all required safety requirements are addressed and implemented. \textit{Implements} dependencies support certification by providing traceability between system requirements and the certification requirements they implement. This traceability also helps verify that system requirements address all certification requirements. Finally, \textit{Conflicts} dependencies help identify conflicting requirements and prevent unresolved conflicts from resulting in contradictions. This is particularly useful for identifying conflicts between functional and safety requirements, which may require reviewing functional requirements against safety constraints. It also facilitates the verification that system requirements do not conflict with one another or, where conflicts are identified, that they do not contradict certification requirements.

Overall, these dependency types provide valuable information that helps maintain the consistency and quality of the SRS as it evolves and supports downstream activities, including software design, implementation, and maintenance.

\section{LEREDD: LLM-Enabled Requirement Dependency Detection Approach}\label{sec:approach}

\subsection{Overview}


LEREDD is an LLM-based dependency detection approach that automatically identifies direct dependencies between pairs of requirements. LEREDD takes an SRS document (containing a list of requirements) and a dataset of annotated requirement pairs as input. The requirement pairs are distinct from the SRS document, and each pair is annotated with its dependency type. 
As output, it predicts the dependency type (or no dependency) for each requirement pair extracted from the SRS, along with a confidence score and a rationale.  

\begin{figure}[htbp]
\centering
\includegraphics[width=\linewidth]{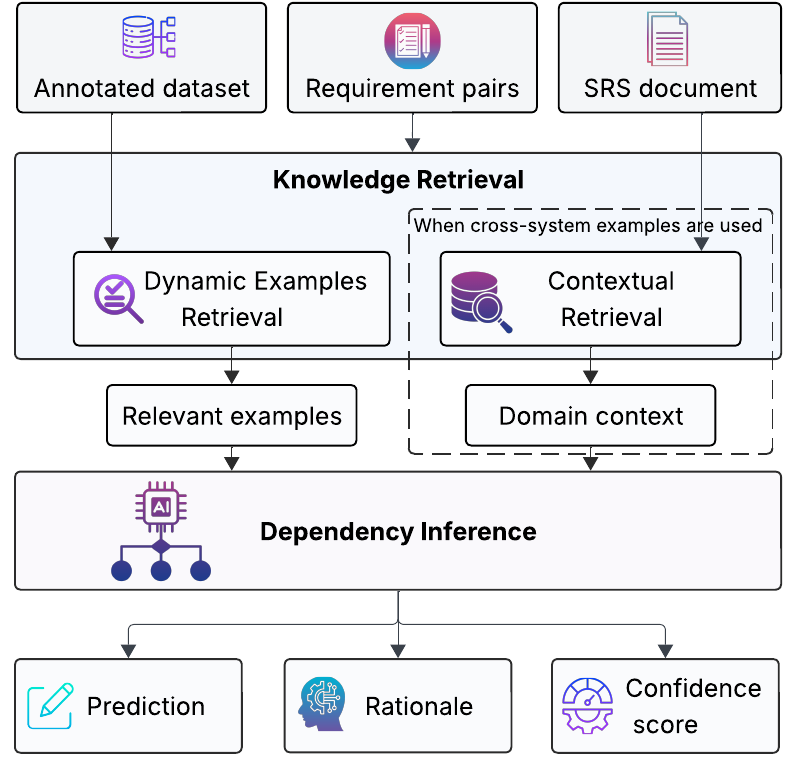}
\caption{LEREDD Framework.}
\label{fig:LEREDDframework}
\end{figure}

The overall framework of LEREDD is illustrated in \autoref{fig:LEREDDframework}. Given a set of $n$ NL requirements extracted from the SRS document, a list of $\frac{n(n-1)}{2}$ requirement pairings is generated, representing the complete set of unique requirement pairs. 
A two-phase pipeline processes these pairs, comprising a \textit{knowledge retrieval} phase followed by a \textit{dependency inference} phase. The \textit{knowledge retrieval} phase applies a dual-strategy process to facilitate in-context learning: (1) \textit{dynamic examples retrieval}, and (2) \textit{contextual retrieval}. The former identifies and retrieves requirement pairs similar to the input pair for all dependency types from the annotated dataset, serving as examples within the prompt.
The \textit{contextual retrieval} employs RAG to extract domain-specific information from the SRS document, which serves as domain-specific context within the prompt. In LEREDD, contextual retrieval is applied only when the few-shot examples originate from a different system than the analyzed requirement pairs, as it was found to reduce performance when the examples originate from the same system.  
The data extracted from that phase informs the \textit{dependency inference} phase, in which an LLM assesses the dependency links between the target requirement pair $(R_1, R_2)$ using the augmented prompt. To enhance the model's reasoning and rigor, the prompt requires the LLM to self-reflect, providing a rationale and a confidence score on a 0–5 rating scale for each prediction~\cite{i2023minimizingfactualinconsistencyhallucination,xu2024sayselfteachingllmsexpress,detommaso2024multicalibrationconfidencescoringllms}. 

\subsection{Knowledge Retrieval}\label{sec:retreival}


The \textit{knowledge retrieval} phase involves extracting contextual information required for the detection task in two stages: \textit{dynamic example retrieval} and \textit{contextual retrieval}. We determined the parameter selection and configuration used in LEREDD for both stages through a systematic analysis supported by an extensive suite of experiments. We present the experimental setup, analysis, and results in Section~\ref{sec:Results}.

\subsubsection{Dynamic examples retrieval}
We employ \textit{dynamic examples retrieval} to facilitate ICL, a paradigm that enhances LLM predictions by augmenting the input prompt with task-specific examples~\cite{dong2024surveyincontextlearning}. Building on existing work (e.g.,~\cite{li2023unifieddemonstrationretrieverincontext,wu2025retrievalaugmentedgenerationnaturallanguage,niu2026tvrautomotiverequirementtraceability}), LEREDD uses dynamic retrieval to select semantically similar examples from the annotated dataset, providing the model with context-specific examples tailored to the target requirement pair. Examples are provided for each dependency type and the no-dependency case to provide a comprehensive context for the classification task.
To support example retrieval, LEREDD leverages the \textit{SBERT} model to generate embeddings for individual requirements in both the target pair $(R_1, R_2)$ and the candidate example pair $(R_a, R_b)$. Semantic similarity is then computed using the Euclidean distance between each requirement in the target pair and its corresponding requirement in the candidate example pair. This process thus yields four scores, which are aggregated by computing their arithmetic mean to obtain a single semantic similarity score for the two pairs. The semantic similarity is calculated according to the following formula:

\begin{equation}
    \scalebox{1}{ 
    $ \text{Score}_{\text{avg}} = \frac{\text{sim}(R_1, R_a) + \text{sim}(R_1, R_b) + \text{sim}(R_2, R_a) + \text{sim}(R_2, R_b)}{4} $
    }
    \label{eq:eq2}
\end{equation}



Next, we retrieve the top six most similar examples to the target requirement pair $(R_1, R_2)$ for each dependency type. We determined the parameter selection and configuration used in LEREDD through a systematic analysis supported by an extensive suite of experiments. Sections~\ref{sec:RQ2Setting} and~\ref{sec:RQ2Results} present the experimental setup, analysis, and results. 

For instance, consider the following target requirement pair: \textit{``The system shall always stop the vehicle to prevent collision with objects during the parking maneuver''} and \textit{``The system shall include the BCS''}.  One of the examples included for the \textit{Requires} dependency is: \textit{``The BCS shall accept camera scenery information provided by the PCS to brake the vehicle if an object is within the vehicle’s path''}, which has a \textit{Requires} dependency with \textit{``The system shall include the BCS''}. This example explicitly states the BCS functionality and establishes the \textit{Requires} dependency between the requirement specifying the BCS's existence and the requirement describing its functionality, both of which support identifying the dependency. In contrast, an example that was not selected because of its low similarity to the target pair is: \textit{``The system shall start identifying available parking spots upon user confirmation''} and \textit{``The system shall allow the customer to select an available parking spot displayed through the HMI with a confirm button''}. Although this example represents a \textit{Requires} dependency, it is unrelated to the BCS and does not establish a dependency involving the existence of a system. We use the aforementioned target pair as a running example throughout this section to illustrate the approach and its outputs.  

Overall, the \textit{dynamic example retrieval} stage produces relevant examples that serve as inputs to the subsequent \textit{dependency inference} stage. 

\subsubsection{Contextual retrieval}
\textit{Contextual retrieval} employs RAG, a technique that enables LLMs to produce context-aware outputs by integrating domain-specific knowledge retrieved from relevant documents~\cite{10.1145/3744746}. RAG is highly relevant to RE, as existing research suggests that its limited adoption constitutes a missed opportunity to leverage LLMs’ full reasoning and retrieval capabilities for RE tasks~\cite{zadenoori2025largelanguagemodelsllms}. In dependency detection, RAG helps identify structural relationships among system components, which can then be mapped to dependencies between their corresponding requirements. 

Within the LEREDD framework, RAG is only employed when few-shot examples originate from a different system. It is applied to the SRS document, focusing on the system description sections and the requirements list. The requirements list is incorporated into the context pool because it provides fundamental information about system architecture and component interactions. A fixed-size chunking strategy is employed in which the six most semantically similar 1000-character chunks, with a 200-character overlap, are retrieved and provided as domain context to the model. This configuration reduces noise and gives the LLM focused context. In the running example, one chunk included in the context contains the following statement: \textit{"Brake Control Subsystem (BCS) – An APA subsystem that brakes the vehicle based on information received from the Park Control Subsystem"}. This statement defines the BCS subsystem and its braking functionality, facilitating identification of the dependency between the two requirements in the running example.

\subsection{Dependency Inference}\label{sec:DepInference}

The dependency inference stage is the core of the LEREDD framework, where the LLM leverages contextual information generated during the retrieval stage to identify dependencies between requirement pairs. For this task, we use the \textit{GPT-5.4} model, which performed best in our evaluations (as reported in \autoref{sec:Results}). 
The LLM prompt includes the top six most similar examples for each dependency type and, when applicable, six chunks of domain-specific context. Beyond contextual information, the prompt includes formal definitions of each dependency type to ensure conceptual clarity and help distinguish types. While LEREDD is extensible to accommodate diverse dependency types, the current prompt supports five types, as described in Section \ref{sec:problem}: \textit{Requires}, \textit{Implements}, \textit{Conflicts}, \textit{Details}, and \textit{Is similar}. 





 


\autoref{fig:prompt} illustrates the prompt employed for inference. The prompt is structured into three primary segments: (1) \textit{contextual scope}, (2) \textit{input data}, and (3) \textit{instructions}. The first segment instructs the model to adopt an expert requirements engineer persona, specifies the domain and system from which the requirements were derived, and defines the detection task. This information grounds the model and provides preliminary orientation for the detection task. The second segment includes the four inputs required for the task: the two requirements to be analyzed, the formal definitions of dependency types, and the examples generated during the \textit{dynamic example retrieval} stage. The dependency definitions explain the concept of dependency to the LLM and the differences among dependency types, which are essential to the task. These inputs provide precise, granular information, thereby facilitating accurate dependency detection. 

\begin{figure*}[ht]
    \centering
    \begin{tcolorbox}[
        colback=boxblue,       
        colframe=gray!80!black, 
        boxrule=1pt,          
        arc=4pt,                
        left=5pt, right=5pt, top=5pt, bottom=5pt 
    ]
    \small
    You are an expert requirements engineer from the \underline{\textit{<domain of interest>}}. You will be provided with a pair of requirements extracted from the software requirements specification for  \underline{\textit{<system name>}}.\\
        Given the following requirement dependency type definitions, examples, and context, your task is to analyze the requirement pair and determine whether a direct or indirect dependency exists between them. \\
        
         \textbf{\#Requirements to analyze:}   \\
        Requirement A: \underline{\textit{<first requirement>}} \\
        Requirement B: \underline{\textit{<second requirement>}} \\
        
        \textbf{\#Dependency Definitions:} \\
        \underline{\textit{<Definitions of dependency types>}} \\
        
        \textbf{\#Examples:}\\
        \underline{\textit{<Retrieved examples>}} \\
        
         \textbf{\#Context:}\\
         \underline{\textit{<Retrieved domain-specific context>}}\\
        
        \textbf{\#Instructions}\\
        - You should analyze the requirements pair according to all dependency types. \\
        - If a dependency exists between a pair, you should annotate it with the type of dependency.\\
        - If it does not fall into one of the above types of dependency, annotate it with ``No\_dependency''. \\
        - Explain the rationale behind your annotation. \\
        - Provide a confidence score for the annotation. The score should range from 0 to 5, with 0 indicating no confidence and 5 indicating the highest confidence. \\
        - **The output MUST be structured using these exact labels, each on a new line:** \\
          **Dependency\_Type: [TYPE]** \\
          **Rationale: [EXPLANATION]** \\
          **Confidence Score: [SCORE]**
    \end{tcolorbox}
    \caption{Dependency Inference Prompt.}
    \label{fig:prompt}
\end{figure*}

The final segment provides specific instructions for detection, expected outputs, and their formats. Specifically, we instruct the model to analyze the requirements according to all dependency types before making a prediction. This ensures the model considers all dependency types before deciding, minimizing the risk of overlooking relevant dependencies. The model is then instructed to annotate the requirement pair with the identified dependency type or with \textit{No\_dependency} if no dependency is detected. The LLM evaluates dependencies in both directions.

Beyond the primary prediction (i.e., the dependency type), the model is instructed to generate a rationale justifying the prediction and a confidence score, measured on a 0–5 rating scale, reflecting its certainty in the generated prediction. These additional outputs are designed to promote self-reflection in the model, thereby reducing the risk of hallucinations. In particular, prompting LLMs to generate rationales enhances their reasoning capabilities while promoting transparency and credibility~\cite{i2023minimizingfactualinconsistencyhallucination,xu2024sayselfteachingllmsexpress}. These rationales provide deeper insight into the LLM's reasoning process, which is particularly useful for identifying issues that may lead to false predictions and for addressing them during subsequent refinement iterations of LEREDD. Furthermore, for directional dependency types, they identify the direction of the dependency, which is essential for correctly interpreting LEREDD's outputs. They are also particularly valuable to requirements engineers who use LEREDD's outputs, as they provide additional context for understanding and assessing the predicted dependencies. Moreover, providing confidence scores improves the interpretability of the generated outputs~\cite{detommaso2024multicalibrationconfidencescoringllms}. These scores also support threshold-based analyses, enabling a filtering mechanism that excludes predictions with low LLM confidence.

Therefore, for each requirement pair, LEREDD generates three outputs: (1) a prediction classifying the dependency type or indicating the absence of a dependency, (2) a confidence score, and (3) a rationale justifying the underlying reasoning. For the running example, the generated outputs are the following: 

\begin{itemize}
    \item \textbf{Dependency\_Type:} Requires
    \item \textbf{Rationale:} Requirement\_A describes a safety function: the vehicle must stop to prevent collision during the parking maneuver. Requirement\_B provides the braking capability needed to achieve that stop by including the Brake Control System (BCS). Therefore, fulfilling Requirement\_A depends on implementing Requirement\_B as a prerequisite. Without the BCS being included to brake the vehicle, the system could not reliably satisfy the stopping behavior in Requirement\_A.
    \item \textbf{Confidence score:} 5.
\end{itemize}

In this example, LEREDD successfully identifies the \textit{Requires} dependency between the two requirements. In its rationale, the LLM follows the expected reasoning by identifying Requirement\_B (i.e., the second requirement regarding the existence of the BCS) as necessary for fulfilling Requirement\_A, which specifies the braking functionality. The rationale also identifies the direction of the Requires dependency, indicating that Requirement\_A requires Requirement\_B. The confidence score of 5 further reflects the LLM's confidence in both its prediction and its rationale.

We developed the prompt following established prompt engineering practices (e.g., \cite{3721041.3721046,schulhoff2025promptreportsystematicsurvey}) and iteratively refined it through multiple rounds of validation, experimentation, and analysis of the LLM's outputs (i.e., rationales) to improve effectiveness. We initially considered multiple prompting techniques (e.g., chain-of-thought prompting and Retrieval-Augmented Generation) but excluded them because they negatively affected the model’s performance on the task.

\section{Empirical Evaluation} \label{sec:empiricalEvaluation}

\subsection{Research Questions}

\textbf{\textit{RQ.1: Which LLM is more effective for requirement dependency detection?}} This RQ aims to empirically evaluate the performance of LLMs 
on requirement dependency detection under a zero-shot prompting setting, and to identify the best-performing model.

\textbf{\textit{RQ.2: What is the optimal configuration for requirement dependency detection?}} This RQ aims to systematically identify the optimal configuration for requirement dependency detection by evaluating different parameter settings for each prompting strategy, including few-shot and RAG.

\textbf{\textit{RQ.3: What are the contributions of contextual retrieval and dynamic example retrieval to LEREDD's performance?}} This RQ investigates the individual and combined contributions of contextual retrieval and dynamic example retrieval to LEREDD's performance using the optimal configurations identified in RQ2.  Additionally, we evaluate the semantic-based example selection strategy adopted in dynamic example retrieval against fixed and random few-shot selection. This RQ provides empirical evidence supporting the design decisions made in LEREDD. 
    
\textbf{\textit{RQ.4: How does LEREDD compare to baseline approaches for intra-dataset dependency detection?}} This RQ aims to evaluate and compare LEREDD with two SOTA baselines, using requirements from the same dataset (system) (1) to select dynamic examples required by LEREDD and (2) to train and test the selected baselines.

\textbf{\textit{RQ.5: How does LEREDD compare to baseline approaches for cross-dataset dependency detection?}} 
This RQ evaluates LEREDD against two baselines in a cross-dataset setting, in which training and evaluation are performed on different datasets. Specifically, we consider two datasets $D_1$ and $D_2$.
For LEREDD, example retrieval is performed using annotated requirement pairs from $D_1$, while the requirement pairs to be classified are drawn from $D_2$. The baseline models are trained on $D_1$ and evaluated on $D_2$. This setup reflects the most common deployment condition, where annotated data are typically available only for previous systems, while the detection is performed on new systems.

\textbf{\textit{RQ.6: To what extent do industrial experts agree with the predictions of LEREDD?}} 
This RQ assesses LEREDD's reliability when evaluated against industrial requirements. Specifically, we measure the degree to which experts from industry agree with the predictions of LEREDD. This RQ assesses the applicability and generalizability of LEREDD to industrial settings.

\subsection{Dataset Construction}

\subsubsection{Data Collection}

To ensure the validity of our experimental data, it must satisfy two criteria: (1) the requirements should be specified in NL and manually annotated for direct dependencies, and (2) the annotated requirements should be recent to ensure that they were not used for training LLMs. To the best of our knowledge, no publicly available dataset meets both criteria.
To mitigate these limitations, we collected publicly available SRS documents from Michigan State University’s Requirements Engineering course website\footnote{\url{https://www.cse.msu.edu/~cse435/\#cinfo}}. These documents were produced within a structured requirements engineering curriculum in collaboration with industrial partners who guided their development and validation. They underwent multiple rounds of feedback and revision, resulting in relatively high-quality specifications that accurately represent industrial practice in regulated domains and are suitable for empirical analysis. In fact, our industrial partner (General Motors) contributed to this course and confirmed that the documents produced through it are both valid and representative of industrial practice. 

From the available materials, we selected three SRS documents corresponding to different automotive systems, each representing a distinct functionality: Traffic Jam Assist (TJA), Automated Parking Assist (APA), and Adaptive Driving Beam (ADB). These systems are widely studied in the automotive domain and exhibit nontrivial interactions among requirements, making them appropriate for dependency analysis. 
We curate the dataset by extracting NL requirements from each SRS and omitting non-requirement sections, such as introductions, abbreviations, and supplementary descriptions. This process yields collections of 40, 25, and 50 requirement statements for the ADB, TJA, and APA systems, respectively.

To evaluate the applicability of LEREDD to industrial requirements, we additionally considered an automotive system from our industrial partner. This system provides an advanced feature for electric vehicles, enabling better control over vehicle acceleration and deceleration. The system comprises 37 requirements. The industrial partner selected it because its requirements have diverse dependencies.    


\subsubsection{Ground Truth} 

To construct the ground truth for requirement dependency analysis, we adopted two different processes, one for the public systems (i.e., ADB, APA, TJA), and another for the industrial system. 

For the public systems, two authors (each with more than five years of experience in requirements engineering) independently annotated the requirement pairs within each system, following a structured, multi-step annotation process.

We integrated the set of requirement dependency types and their formal definitions from Section \ref{sec:DepInference} and augmented them with representative examples extracted from the handbook~\cite{Motger2025}. This formed the annotation guidelines. All researchers involved in the project developed and agreed on the annotation guidelines, and our industrial partner reviewed and validated them. The two annotators then independently analyzed and annotated requirement pairs within each system according to these guidelines. Both annotators thoroughly reviewed the SRS documents (including system descriptions and design) to acquire the domain knowledge required for the annotation task. Each requirement pair was labeled with a single dependency type or marked as having no dependency. For pairs with multiple dependency types, we retained only the most evident dependency. Because the annotation process is manual, it is both costly and time-consuming, even for SRS documents containing a relatively small number of requirements. For instance, the ABD document comprises 40 requirements, yielding 780 possible unique requirement pairs. Exhaustively annotating all such pairs is therefore impractical. 

To prioritize annotation effort on requirement pairs more likely to be dependent, we embedded each requirement using all-MiniLM-L6-v2, a pretrained BERT-based sentence encoder, and computed cosine similarity between all requirement pairs. We selected all-MiniLM-L6-v2 due to its demonstrated efficiency and effectiveness in capturing semantic similarity~\cite{reimers2019sentencebertsentenceembeddingsusing}. We then ranked the pairs in descending order of similarity. We then manually annotated the pairs, starting with the most semantically similar requirement pairs. This similarity-based ranking was performed strictly to reduce manual annotation effort and is not part of the LEREDD approach. Randomly sampling pairs would likely have resulted in a lower proportion of dependent pairs, while still requiring considerable manual annotation effort. For the ADB system, we stopped annotation after 413 requirement pairs because the number of dependent pairs had plateaued. From this dataset, we used a randomly selected subset of 100 requirement pairs to identify the optimal configuration for LEREDD (i.e., to address RQ2), while the remaining 313 pairs were used for the evaluations in RQ3–RQ5. For the TJA and APA systems, we annotated the top 200 most similar requirement pairs for each system. This resulted in 813 annotated requirement pairs across the three systems.

The manual annotation process was highly time-consuming, requiring multiple rounds of validation and three weeks to establish reliable ground truth. This highlights the high cost of manual dependency detection, which LEREDD aims to address. 
We resolved all disagreements by consensus, resulting in a unified set of annotations that serves as the gold standard for evaluation. 

For the industrial system, we followed a different evaluation process to address RQ6, i.e., to assess agreement between industrial experts and LEREDD's predictions. The system comprised 37 requirements, resulting in 666 unique requirement pairs. We first ran LEREDD on all 666 pairs. Since it was impractical for the industrial experts to manually review all pairs, we aimed to minimize their annotation effort. To this end, we first selected all 76 requirement pairs that LEREDD identified as dependent. We then randomly selected 40 non-dependent pairs from LEREDD's predictions. This resulted in 116 requirement pairs for the industrial case study, which we then provided to the industrial experts.
To evaluate the correctness of LEREDD's predictions for these pairs, three experts from our industrial partner manually reviewed the generated predictions, rationales, and confidence scores. For each pair, they indicated whether they agreed with LEREDD's prediction or identified a different dependency type or no dependency. These experts were familiar with the system under study and therefore had the domain knowledge required for this task. They received the same annotation guidelines, including definitions and examples of the dependency types, to guide their review. All disagreements were discussed, and we used majority voting to reach consensus.

After independent annotation on all four systems, we measured inter-annotator agreement using Cohen’s kappa score to assess consistency beyond chance. The overall Cohen's kappa was 0.50, indicating moderate agreement, reflecting the task's inherent subjectivity and complexity. Furthermore, two requirements can have more than one type of dependency between them. However, for cost-efficiency, we labeled only the most evident dependency type. This constraint may also have contributed to disagreements among annotators, as they may have differed in their judgments of which dependency type was the most evident. This is expected to worsen our accuracy results, as LEREDD can identify a correct dependency type that is nonetheless counted as incorrect because it differs from the one labeled in the ground truth.

Table~\ref {tab:dataset} summarizes the distributions of dependency types across systems. Across all datasets, \textit{No dependency} represents the majority of requirement pairs, accounting for more than 71\% of all pairs. The \textit{Requires} dependency is the most frequent dependency type, accounting for 13\%, 9\%, and 23\% of requirement pairs in ADB, TJA, and APA, respectively. It is followed by the \textit{Implements} dependency, while the other dependency types are scarce across all datasets. This class imbalance reflects the inherent nature of requirement dependencies rather than a characteristic specific to our datasets. Our industrial partners confirmed this observation based on their experience with their systems. It is also consistent with prior work on requirement dependency detection. For instance, an analysis of the PURE dataset found that only 11.5\% of requirement pairs were related, while the remaining 88.5\% were unrelated \cite{Motger2025}.  
We provide the annotated dataset in our replication package. 

The subset used to determine the optimal LEREDD configuration comprises 100 requirement pairs, including 13 pairs exhibiting a \textit{requires} dependency, 6 \textit{implements} dependencies, 2 \textit{details} dependencies, 1 \textit{conflicts} dependency, and 78 pairs with \textit{No\_dependency}. Accordingly, we considered only these dependency types in the experiments.

\begin{table*}[t]
\centering
\caption{Distribution of Requirement Dependency Types}
\label{tab:dataset}
\begin{tabular}{l|p{1.23cm}|p{1.1cm}|l|l|c|l}
\hline
\multirow{2}{*}{\textbf{Dependency Type}} & \multicolumn{2}{c|}{\textbf{ADB}} & \multirow{2}{*}{\textbf{TJA}} & \multirow{2}{*}{\textbf{APA}} & \multirow{2}{*}{\textbf{Industrial system}} & \multirow{2}{*}{\textbf{Total}} \\ \cline{2-3}

 & Pilot study & Evaluation &  &  & & \\
\hline
\# Requirement pairs & 100 & 313& 200 & 200 & 116 &929 \\
\hline
Conflicts      & 1 & 10& - & 4 & 1 &16 \\
Details        & 2 & 13 & 2 & 1 & 19 &37 \\
Implements     & 6 & 18 & 10 & 3 & 30 &67 \\
Is similar     & - & 3& 1 & 3 & - &7 \\
Requires       & 13 & 40& 17 & 45 & 26 &141 \\
No Dependency  & 78 & 229& 170 & 144 & 40 &661 \\
\hline
\end{tabular}
\end{table*}

\subsection{LLMs Selection}

To evaluate the effectiveness of SOTA LLMs in identifying requirement dependencies, we select the models based on the following criteria: (1) they should demonstrate strong performance on RE-related tasks, (2) they should originate from different model families, and (3) they should include both proprietary and open-source models. These criteria aim to enhance the diversity of the selected models. As a result, we select four representative models: \textit{GPT-5.4}, \textit{Llama 3.1}, \textit{Gemma 12B}, and \textit{Mistral-nemo (12B)}. These models have demonstrated strong performance in RE research~\cite{zadenoori2025largelanguagemodelsllms, 10.1145/3695988}; they belong to distinct model families: OpenAI GPT, Meta Llama, Google Gemma, and Mistral AI, and they collectively cover both proprietary and open-source models. The prompt shown in~\autoref{fig:prompt} serves as the basis for all evaluation techniques, with variations in the context added to the prompt: no examples or additional context for zero-shot, and added examples for few-shot. 

\subsection{Parameter Settings for Configuration Selection}\label{sec:RQ2Setting}

To establish the optimal configuration for requirement dependency detection for RQ.2, we investigate multiple parameter settings for: (1) few-shot prompting to support the \textit{dynamic example retrieval} stage, and (2) Retrieval-Augmented Generation (RAG) as a complementary technique to few-shot prompting, a technique that enables LLMs to produce context-aware outputs by integrating domain-specific knowledge retrieved from relevant documents~\cite{10.1145/3744746}. We chose RAG because it is highly relevant to RE; existing research suggests that its limited adoption is a missed opportunity to leverage LLMs’ full reasoning and retrieval capabilities for RE tasks~\cite{zadenoori2025largelanguagemodelsllms}. In dependency detection, RAG can help identify structural relationships among system components, which can then be mapped to dependencies between their corresponding requirements. For instance, given the requirements: \textit{``The system shall always stop the vehicle to prevent collision with objects during the parking maneuver''} and \textit{``The system shall include the BCS''}, a dependency is only apparent when the context (i.e., defining the BCS as the subsystem responsible for braking) is provided. 


\subsubsection{Few-Shot Prompting}

Regarding few-shot prompting, we investigated four parameters: (1) the embedding model, (2) the similarity metric, (3) the formula for computing similarity between requirement pairs, and (4) the number of examples (\textit{k}). For embedding generation, we tested Sentence-BERT (SBERT), specifically the all-mpnet-base-v2~\footnote{\url{https://www.sbert.net/docs/sentence_transformer/pretrained_models.html}} 
model, and BGE-M3~\footnote{\url{https://bge-model.com/bge/bge_m3.html}} 
(BAAI General Embedding) for their proven efficacy in capturing sentence-level semantic similarity. We measure similarity between embeddings using cosine similarity or Euclidean distance, as they remain standard, widely used measures for such tasks. To identify the optimal method for calculating similarity between the target requirement pair $(R_1, R_2)$ and the candidate example pair $(R_a, R_b)$, we compared two aggregation formulas. The first formula \eqref{eq:eq2} (introduced in Section~\ref{sec:retreival}) computes the arithmetic mean of the similarity scores of all four requirement combinations.
The second formula \eqref{eq:eq1} averages the maximum similarity scores associated with each requirement in the target pair. 
 We also varied the number of examples per dependency type in the prompt from 1 to 9.

\begin{equation}
    \scalebox{0.91}{ 
    $ \text{Score}_{\text{max\_avg}} = \frac{\max(\text{sim}(R_1, R_a), \text{sim}(R_1, R_b)) + \max(\text{sim}(R_2, R_a), \text{sim}(R_2, R_b))}{2} $
    }
    \label{eq:eq1}
\end{equation}

With the selected few-shot parameters and variations, we identified \textbf{72 configurations}. To ensure experimental robustness and result reliability, we ran each configuration three times at a temperature of 0.2. Lower temperature settings generally suit tasks requiring strong instruction-following and stable outputs \cite{LI2025242}. Accordingly, we selected a temperature of 0.2 to support stable, consistent dependency detection while preserving limited variability across runs. Overall, we conducted \textbf{216 experiments} for few-shot. 

Furthermore, we leveraged the confidence scores to conduct a threshold-based analysis. Specifically, requirement pairs assigned a confidence score of 4 or lower were automatically reclassified as \textit{No\_dependency}. We selected the threshold of 4 based on preliminary exploratory analysis, which indicated that the model rarely produced confidence scores below this value. Each experiment was evaluated both with and without threshold-based re-annotation, resulting in a total of \textbf{432 evaluations}.

\subsubsection{Retrieval Augmented Generation}

In the evaluation, we apply RAG to the SRS document, focusing on the system description sections and the requirements list. We incorporate the requirements list into the context pool because it provides fundamental information about system architecture and component interactions. We use a fixed-size chunking strategy, retrieving the $k$ most semantically similar chunks and providing them as domain context to the model.
For the experiments, we investigated two parameters: (1) chunk size, experimenting with 500 and 1000 characters per chunk with 200 character overlap, and (2) the number of chunks included in the prompt, considering 2, 6, and 10 chunks, as well as the full document. Accordingly, \textbf{seven configurations} were identified, resulting in 21 experiments in total, with each configuration executed three times. We conducted all RAG experiments using the optimal configuration identified in the few-shot prompting experiments. 

\subsection{Baselines}

\subsubsection{Selection Criteria}

To evaluate the LEREDD framework, we selected baselines based on four main criteria. First, the baselines should span different categories of approaches to ensure diversity and to support rigorous, representative comparisons with the broader research landscape. Second, these approaches should be among the most prevalent and well-established within their respective categories. This criterion facilitates fair comparison across categories. Third, the approaches should be applicable across different systems and domains. 
This criterion is essential because our evaluation encompasses multiple systems. Finally, baselines should provide either open-source implementations or comprehensive configuration guidelines to ensure reproducible experimental results.  

\subsubsection{Selected Baselines}

We excluded existing LLM-based methods because they either target specific application contexts outside the scope of this study (e.g., approaches based on formal logic or applied to mobile app reviews) or focus on dependency types underrepresented in our datasets. Specifically, G\"{a}rtner~\textit{et al.}~\cite{Gaertner2024ASE} exclusively target contradiction dependencies, whereas our datasets do not contain such dependencies. Therefore, we cannot evaluate this approach on our datasets. Although Almoqren~\textit{et al.}~\cite{app15189891} support multiple dependency types, it requires a knowledge graph, which can be costly to construct. Furthermore, this approach is specifically designed for mobile app reviews. Finally, Niu~\textit{et al.}~\cite{niu2026tvrautomotiverequirementtraceability} focus on identifying trace links between stakeholder requirements and system requirements, rather than between system requirements, which is the focus of our study.  
 
Based on the first and second criteria, we considered three approaches: (1) a retrieval-based approach using \textit{TF-IDF \& LSA}~\cite{8995268}, (2) a knowledge-based approach using \textit{ontologies}~\cite{Motger2019OpenReqDDAR}, and (3) an ML-based approach using \textit{fine-tuned BERT}~\cite{Graler_Oleff_Hieb_Preuss_2022}. The selected approaches ensure diversity and represent prevalent approaches within their respective categories. TF-IDF is a fundamental lexical method, and its integration with LSA provides a strong baseline for lexical and semantic retrieval~\cite{Motger2025,8995268}. Ontologies provide more effective domain modeling compared to graphs~\cite{Motger2025}. Fine-tuned BERT is a prominent ML approach known for its accuracy in multi-class classification tasks and outperforms conventional approaches~\cite{Prabhu2021MulticlassTC,10.1145/3442442.3451375,Graler_Oleff_Hieb_Preuss_2022}. Based on the third criterion, we excluded the ontology-based approach. Although it may offer high precision through domain-specific representation, we omitted it because ontology development is costly and scalability is limited. Finally, based on the fourth criterion, we retained the \textit{TF-IDF \& LSA} and \textit{fine-tuned BERT} baselines, as the former is supported by standard Python libraries and the latter by open-source implementations. 

To evaluate LEREDD against the selected baselines, we implemented \textit{TF-IDF \& LSA} following the configuration described by Samer~\textit{et al.}~\cite{8995268}, where it serves as a recommender for \textit{Requires} dependencies. For fine-tuned BERT, we adhered to the configuration guidelines described by Gräßler~\textit{et al.}~\cite{Graler_Oleff_Hieb_Preuss_2022}. Their approach detects \textit{Requires} and \textit{Refines} dependencies and uses oversampling and class weighting to address class imbalance. We extended this configuration to cover additional dependency classes (e.g., \textit{Implements} and \textit{Conflicts}) by adding new classes and including corresponding instances in the training dataset.
To ensure a fair comparison, we systematically tuned hyperparameters for both baselines on the same ADB subset used for the pilot study to identify their optimal configurations. For LEREDD and fine-tuned BERT, we opted for an 80\%/20\% split of the ground truth for intra-dataset experiments. For BERT, we used the 80\% for training, consistent with the original approach~\cite{Graler_Oleff_Hieb_Preuss_2022}, while for LEREDD, it served as the example pool for few-shot retrieval. Note that this 80\%/20\% split is not required for LEREDD, but we adopted it to enable a fair comparison with the fine-tuned BERT baseline.

\section{Empirical Results}\label{sec:Results}

\subsection{\textbf{RQ1. Performance of SOTA LLMs for requirement dependency detection}}

To assess the effectiveness of SOTA LLMs in identifying requirement dependencies, we evaluate four models: \textit{GPT-5.4}, \textit{Llama 3.1}, \textit{Gemma 12B}, and \textit{Mistral-nemo} using a zero-shot setting, without considering any in-context examples or task-specific knowledge. 
Table~\ref{tab:zeroShotResults} reports the zero-shot performance of the four LLMs on the three automotive system datasets: D1 (ADB), D2 (TJA), and D3 (APA). Throughout the paper, the overall precision, recall, and $F_1$-score refer to weighted metrics, which account for the distribution of dependency types in each dataset. This ensures that the overall results reflect their actual distribution. However, since performance across individual dependency types is important, we additionally report class-level precision, recall, and $F_1$-scores for each dependency type.  

As Table~\ref{tab:zeroShotResults} shows, GPT-5.4 consistently outperforms all other evaluated models. It achieves the highest $F_1$-scores across all datasets (D1: 0.78, D2: 0.84, D3: 0.71), with an average of 0.78, indicating stronger zero-shot dependency classification generalization than other models. It also achieves the highest accuracy across all datasets, ranging from 0.66 for D3 to 0.82 for D2. In contrast, Llama 3.1, Gemma 12B, and Mistral-nemo show lower and more variable performance, suggesting that dependency detection remains challenging for most open-source models. Among the open-source models, Llama achieves the highest $F_1$-scores, averaging 0.69, followed closely by Gemma at 0.65. Llama also achieves the accuracy closest to GPT-5.4 (D1: 0.64, D2: 0.81, D3: 0.56), particularly on D2. However, its $F_1$-scores remain low because it performs poorly on minority dependency types. Mistral remains the lowest-performing model, with an average $F_1$-score of 0.48 across all datasets. Across all models, average precision and recall are relatively balanced, indicating that no model consistently favors precision over recall, or vice versa.

\begin{table*}[t]
\centering
\scriptsize
\caption{Experimental Results of Zero-Shot LLMs}
\label{tab:zeroShotResults}
\resizebox{\textwidth}{!}{
\begin{tabular}{c|c|c|ccc|c|ccc|c|ccc}
\hline
\multirow{2}{*}{\textbf{LLM}}  & \textbf{Dependency} & \multicolumn{4}{c|}{\textbf{D1}}  & \multicolumn{4}{c|}{\textbf{D2}} & \multicolumn{4}{c}{\textbf{D3}}  \\ \cline{3-14}
 & \textbf{Type} & \textbf{Acc} & \textbf{P} & \textbf{R} & \textbf{F1} & \textbf{Acc} & \textbf{P} & \textbf{R} & \textbf{F1} & \textbf{Acc} & \textbf{P} & \textbf{R} & \textbf{F1} \\ \hline
\multirow{7}{*}{\textit{GPT-5.4}}& No Dependency & \multirow{7}{*}{\textbf{0.78} $\pm$ 0.01} & 0.89 $\pm$ 0.01 & 0.88 $\pm$ 0.00 & 0.88 $\pm$ 0.00& \multirow{7}{*}{\textbf{0.82} $\pm$ 0.00} & 0.96 $\pm$ 0.00&0.88 $\pm$ 0.00&0.92 $\pm$ 0.00& \multirow{7}{*}{\textbf{0.66}$\pm$0.01} & 0.93 $\pm$ 0.00& 0.71 $\pm$ 0.02& 0.80 $\pm$ 0.01\\
& Requires &  & 0.36 $\pm$ 0.01& 0.44 $\pm$ 0.03 & 0.40 $\pm$ 0.02&  &0.39 $\pm$ 0.04&0.61 $\pm$ 0.06&0.47 $\pm$ 0.05&  & 0.46 $\pm$0.02 & 0.62 $\pm$ 0.04& 0.53 $\pm$ 0.01\\
& Implements &  & 0.57 $\pm$ 0.08& 0.64 $\pm$ 0.07&0.60 $\pm$ 0.08&  &0.33 $\pm$ 0.05&0.40 $\pm$ 0.00&0.36 $\pm$ 0.03&  & 0.07 $\pm$ 0.01 & 0.56 $\pm$ 0.16&0.13 $\pm$ 0.03\\
& Conflicts &  & 0.47 $\pm$ 0.05 &0.22 $\pm$ 0.00&0.30 $\pm$ 0.01&  & N/A & N/A & N/A &  &0.67 $\pm$ 0.47&0.17 $\pm$ 0.12&0.27 $\pm$ 0.19\\
& Details&  & 0.25 $\pm$ 0.00&0.20 $\pm$ 0.00&0.22 $\pm$ 0.00&  & 0.00 $\pm$ 0.00& 0.00 $\pm$ 0.00&0.00 $\pm$ 0.00&  &0.00 $\pm$ 0.00&0.00 $\pm$ 0.00&0.00 $\pm$ 0.00\\
& Is similar &  &0.00 $\pm$ 0.00&0.00 $\pm$ 0.00&0.00 $\pm$ 0.00&  &0.50 $\pm$ 0.00&1.00 $\pm$ 0.00&0.67 $\pm$ 0.00&  &0.11 $\pm$ 0.16& 0.11 $\pm$ 0.16 & 0.11 $\pm$ 0.16 \\
\cline{2-2} \cline{4-6} \cline{8-10} \cline{12-14}
& \textit{Overall} &  & 0.78 $\pm$0.02& \textbf{0.78} $\pm$ 0.01& \textbf{0.78} $\pm$ 0.01&  & \textbf{0.86} $\pm$ 0.00&\textbf{0.82} $\pm$ 0.00&\textbf{0.84} $\pm$ 0.00&  & \textbf{0.79} $\pm$ 0.02&\textbf{0.67}  $\pm$ 0.02& \textbf{0.71}$\pm$ 0.02 \\
\hline
\multirow{7}{*}{\textit{Llama 3.1}}& No Dependency & \multirow{7}{*}{0.64$\pm$0.00} &0.86 $\pm$  0.00&0.72 $\pm$ 0.01&0.79 $\pm$ 0.00& \multirow{7}{*}{0.81$\pm$0.01} &0.96 $\pm$ 0.00&0.87 $\pm$ 0.01&0.92 $\pm$ 0.00& \multirow{7}{*}{0.56$\pm$0.00} &0.89 $\pm$ 0.01&0.51  $\pm$ 0.00&0.65  $\pm$ 0.00\\
& Requires &  &0.16 $\pm$ 0.00&0.41 $\pm$ 0.01&0.24 $\pm$ 0.00&  &0.36 $\pm$ 0.02&0.82 $\pm$ 0.00&0.50 $\pm$ 0.02&  &0.33 $\pm$ 0.00&0.82 $\pm$ 0.02&0.47 $\pm$ 0.01\\
& Implements &  &0.00 $\pm$ 0.00&0.00 $\pm$ 0.00&0.00 $\pm$ 0.00&  &0.00 $\pm$ 0.00&0.00 $\pm$ 0.00&0.00 $\pm$ 0.00&  &0.00 $\pm$ 0.00&0.00 $\pm$ 0.00&0.00 $\pm$ 0.00\\
& Conflicts &  &0.00 $\pm$ 0.00&0.00 $\pm$ 0.00&0.00 $\pm$ 0.00&  & N/A & N/A & N/A &  & 0.18 $\pm$ 0.02&0.25 $\pm$ 0.00&0.21 $\pm$ 0.01\\
& Details &  &0.00 $\pm$ 0.00&0.00 $\pm$ 0.00&0.00 $\pm$ 0.00&  &0.00 $\pm$ 0.00&0.00 $\pm$ 0.00&0.00 $\pm$ 0.00&  &0.00 $\pm$ 0.00&0.00 $\pm$ 0.00&0.00 $\pm$ 0.00\\
& Is similar &  &0.00 $\pm$ 0.00&0.00 $\pm$ 0.00&0.00 $\pm$ 0.00&  &0.00 $\pm$ 0.00&0.00 $\pm$ 0.00&0.00 $\pm$ 0.00&  &0.00 $\pm$ 0.00&0.00 $\pm$ 0.00&0.00 $\pm$ 0.00\\
\cline{2-2} \cline{4-6} \cline{8-10} \cline{12-14}
& \textit{Overall} &  &0.72 $\pm$ 0.01&0.64 $\pm$ 0.01&0.67 $\pm$ 0.01&  &0.85 $\pm$ 0.00&0.81 $\pm$ 0.01&0.82 $\pm$ 0.01&  &0.72 $\pm$ 0.01&0.56 $\pm$ 0.01&0.58 $\pm$ 0.01\\
\hline
\multirow{7}{*}{Gemma 12B}& No Dependency & \multirow{7}{*}{0.61$\pm$0.00} &0.92 $\pm$  0.00&0.66 $\pm$ 0.00&0.77 $\pm$ 0.00& \multirow{7}{*}{0.67$\pm$0.00} &0.97 $\pm$ 0.00&0.70 $\pm$ 0.00&0.81 $\pm$ 0.00& \multirow{7}{*}{0.54$\pm$0.00} &0.93 $\pm$ 0.01&0.52  $\pm$ 0.00&0.66  $\pm$ 0.00\\
& Requires &  &0.20 $\pm$ 0.00&0.68 $\pm$ 0.00&0.31 $\pm$ 0.00&  &0.19 $\pm$ 0.00&0.82 $\pm$ 0.00&0.30 $\pm$ 0.00&  &0.30 $\pm$ 0.00&0.70 $\pm$ 0.01&0.42 $\pm$ 0.00\\
& Implements &  &1.00 $\pm$ 0.00&0.09 $\pm$ 0.00&0.17 $\pm$ 0.00&  &0.00 $\pm$ 0.00&0.00 $\pm$ 0.00&0.00 $\pm$ 0.00&  &0.10 $\pm$ 0.00&0.33 $\pm$ 0.00&0.15 $\pm$ 0.01\\
& Conflicts &  &0.38 $\pm$ 0.00&0.33 $\pm$ 0.00&0.35 $\pm$ 0.00&  & N/A & N/A & N/A &  &  0.00 $\pm$  0.00& 0.00 $\pm$  0.00& 0.00 $\pm$  0.00\\
& Details &  &0.00 $\pm$ 0.00&0.00 $\pm$ 0.00&0.00 $\pm$ 0.00&  &0.00 $\pm$ 0.00&0.00 $\pm$ 0.00&0.00 $\pm$ 0.00&  & 0.00 $\pm$  0.00& 0.00 $\pm$  0.00& 0.00 $\pm$  0.00\\
& Is similar &  &0.00 $\pm$ 0.00&0.00 $\pm$ 0.00&0.00 $\pm$ 0.00&  &0.00 $\pm$ 0.00&0.00 $\pm$ 0.00&0.00 $\pm$ 0.00&  & 0.00 $\pm$  0.00& 0.00 $\pm$  0.00& 0.00 $\pm$  0.00\\
\cline{2-2} \cline{4-6} \cline{8-10} \cline{12-14}
& \textit{Overall} &  &\textbf{0.80} $\pm$ 0.00&0.61 $\pm$ 0.01&0.66 $\pm$ 0.00&  &0.84 $\pm$ 0.00&0.67 $\pm$ 0.01&0.72 $\pm$ 0.01&  &0.74 $\pm$ 0.01&0.54 $\pm$ 0.01&0.57 $\pm$ 0.01\\
\hline
\multirow{7}{*}{\textit{Mistral-nemo}}& No Dependency & \multirow{7}{*}{0.50$\pm$0.01} &0.93 $\pm$ 0.02 &0.48 $\pm$ 0.01&0.63 $\pm$ 0.01& \multirow{7}{*}{0.52$\pm$0.01} &0.90 $\pm$ 0.01&0.55 $\pm$ 0.01&0.68 $\pm$ 0.02& \multirow{7}{*}{0.42$\pm$0.01} &0.74 $\pm$ 0.01&0.41  $\pm$ 0.01&0.52  $\pm$ 0.01\\
& Requires &  &0.19 $\pm$ 0.01&0.90 $\pm$ 0.01&0.31 $\pm$ 0.01&  &0.12 $\pm$ 0.02&0.67 $\pm$ 0.11&0.20 $\pm$ 0.03&  &0.24 $\pm$ 0.01&0.58 $\pm$ 0.05 &0.34 $\pm$ 0.01\\
& Implements &  &0.48 $\pm$ 0.03&0.33 $\pm$ 0.04&0.39 $\pm$ 0.03&  &0.00 $\pm$ 0.00&0.00 $\pm$ 0.00&0.00 $\pm$ 0.00&  &0.02 $\pm$ 0.03&0.11 $\pm$ 0.16&0.04 $\pm$ 0.05\\
& Conflicts &  &0.00 $\pm$ 0.00&0.00 $\pm$ 0.00&0.00 $\pm$ 0.00&  & N/A & N/A & N/A &  & 0.00 $\pm$ 0.00&0.00 $\pm$ 0.00&0.00 $\pm$ 0.00\\
& Details &  &0.33 $\pm$ 0.47&0.07 $\pm$ 0.09&0.11 $\pm$ 0.16&  &0.00 $\pm$ 0.00&0.00 $\pm$ 0.00&0.00 $\pm$ 0.00&  &0.00 $\pm$ 0.00&0.00 $\pm$ 0.00&0.00 $\pm$ 0.00\\
& Is similar&  &0.00 $\pm$ 0.00&0.00 $\pm$ 0.00&0.00 $\pm$ 0.00&  &0.00 $\pm$ 0.00&0.00 $\pm$ 0.00&0.00 $\pm$ 0.00&  &0.00 $\pm$ 0.00&0.00 $\pm$ 0.00&0.00 $\pm$ 0.00\\
\cline{2-2} \cline{4-6} \cline{8-10} \cline{12-14}
& \textit{Overall} &  &0.78 $\pm$ 0.03&0.50 $\pm$ 0.01&0.56 $\pm$ 0.02&  & 0.77 $\pm$ 0.02&0.52 $\pm$ 0.02&0.59 $\pm$ 0.02&  &0.59 $\pm$ 0.02&0.42 $\pm$ 0.01&0.45 $\pm$ 0.02\\
\hline
\end{tabular}}
\end{table*}

Across all models, the \textit{No Dependency} class consistently achieves the highest performance. GPT-5.4 obtains $F_1$-scores between 0.80 and 0.92 across datasets, averaging 0.87, while the best-performing open-source model, Llama, averages 0.79 for this class. These results suggest that LLMs find it relatively easy to distinguish unrelated requirement pairs, likely because this class has clearer semantic separation. In contrast, models struggle with specific dependency types, often obtaining an $F_1$-score of 0.00 for these classes and averaging 0.29 across dependency types and datasets. For instance, while GPT-5.4 achieves $F_1$-scores ranging from 0.11 to 0.6 for most dependency types across datasets, it scores 0.00 for \textit{Is Similar} in D1 and \textit{Details} in D2 and D3. This is likely due to the extremely limited number of instances for these classes: three for \textit{Is Similar} in D1 and two and one for \textit{Details} in D2 and D3, respectively. Such sparse class distributions make the $F_1$-score highly sensitive to false predictions. The open-source models are much more sensitive to this class imbalance. For instance, Llama achieves an $F_1$-score of 0.00 for 10 dependency types across datasets, while Gemma and Mistral each do so for eight dependency types. These findings suggest that identifying specific dependency types requires deeper semantic understanding and domain knowledge, posing challenges for zero-shot LLMs without task-specific guidance. Overall, GPT-5.4's superior performance led to its selection as the core engine for LEREDD.

\begin{formal}
\textbf{Answer to RQ1:} While GPT-5.4 tends to outperform other models, zero-shot models typically over-predict dependencies. Zero-shot GPT-5.4 can reliably identify non-dependent requirement pairs, achieving 
an average $F_1$-score of 0.87 for the \textit{No dependency} class across the three datasets. However, the model struggles with specific dependency types, yielding an average 
$F_1$-score of only 0.29, highlighting the limitations of zero-shot dependency understanding. 
\end{formal}

\subsection{\textbf{RQ2. Optimal Prompt Configuration for Requirement Dependency Detection}}\label{sec:RQ2Results}

To establish the optimal configuration for requirement dependency detection, we conducted a pilot study using 100 requirement pairs from the ADB (D1) dataset. We investigated multiple few-shot prompting parameter settings to support the \textit{dynamic example retrieval} stage. In addition, we evaluated RAG as a complementary technique to few-shot prompting by assessing various parameter settings. GPT-5.4 was used in the experiments because it was identified as the best-performing LLM based on RQ1. 


\subsubsection{Few-Shot Prompting}

\begin{figure}[htbp]
\centering
\includegraphics[width=\linewidth]{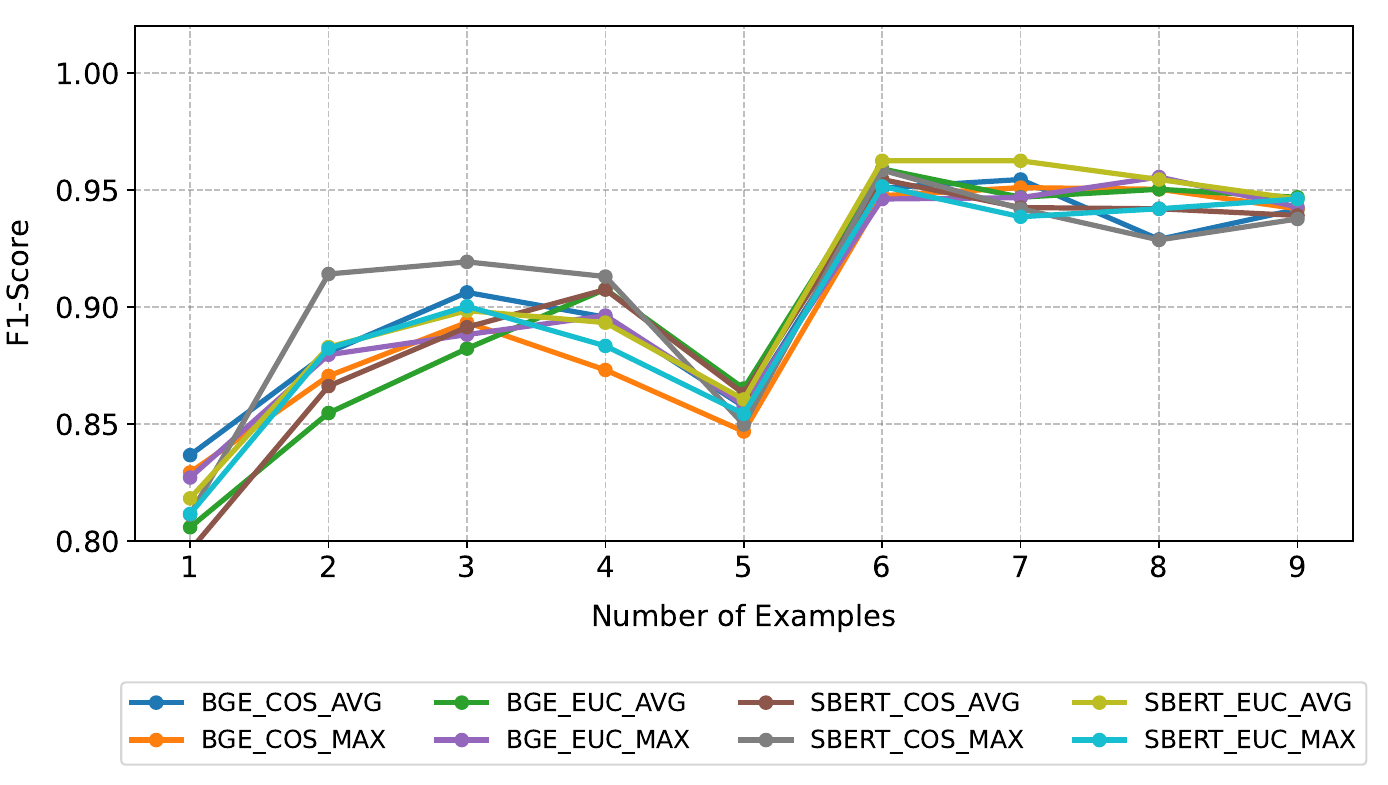}
\caption{$F_1$-score Comparison of Few-shot Configurations}
\label{fig:FewShotConfigurations}
\end{figure}

We evaluated performance using the overall $F_1$-score across the three experimental runs, along with the corresponding standard deviation. For each configuration, we computed the average $F_1$-score with and without threshold-based re-annotation. Comparing these variants showed that threshold-based re-annotation consistently performed better, yielding a higher $F_1$-score in 68\% of configurations (49 out of 72). It also achieved the highest overall $F_1$-score of 0.96. \autoref{tab:fewShotWithThreshold} summarizes the overall $F_1$-scores and standard deviations for all configurations with threshold-based re-annotation, while \autoref{fig:FewShotConfigurations} compares the corresponding $F_1$-scores across the different few-shot configurations under the same setting. We provide the results without threshold-based re-annotation in the replication package. 

\begin{table*}[!htbp]
\centering
\caption{Results of Few-shot Prompting for Configuration Selection with threshold-based re-annotation}
\label{tab:fewShotWithThreshold}

\begin{tabular}{l|cc|cc|cc|cc}
\hline

\multirow{3}{*}{\textbf{Examples}} & \multicolumn{4}{c|}{\textbf{BGE}} & \multicolumn{4}{c}{\textbf{SBERT}} \\ \cline{2-9}

& \multicolumn{2}{c|}{\textbf{Cosine}} & \multicolumn{2}{c|}{\textbf{Euclidean}} & \multicolumn{2}{c|}{\textbf{Cosine}} & \multicolumn{2}{c}{\textbf{Euclidean}} \\ \cline{2-9}

& \textbf{Avg} & \textbf{Max} & \textbf{Avg} & \textbf{Max} & \textbf{Avg} & \textbf{Max} & \textbf{Avg} & \textbf{Max} \\ \hline

\textbf{1} & 0.84 $\pm$ 0.01 & 0.83 $\pm$ 0.02 & 0.81 $\pm$ 0.01 & 0.83 $\pm$ 0.00 & 0.80 $\pm$ 0.01 & 0.81 $\pm$ 0.01 & 0.82 $\pm$ 0.02 & 0.81 $\pm$ 0.02 \\ \hline

\textbf{2} & 0.88 $\pm$ 0.01 & 0.87 $\pm$ 0.01 & 0.85 $\pm$ 0.01 & 0.88 $\pm$ 0.01 & 0.87 $\pm$ 0.01 & 0.91 $\pm$ 0.01 & 0.88 $\pm$ 0.02 & 0.88 $\pm$ 0.01 \\ \hline

\textbf{3} & 0.91 $\pm$ 0.01 & 0.89 $\pm$ 0.01 & 0.88 $\pm$ 0.02 & 0.89 $\pm$ 0.00 & 0.89 $\pm$ 0.00 & 0.92 $\pm$ 0.01 & 0.90 $\pm$ 0.00 & 0.90 $\pm$ 0.02 \\ \hline

\textbf{4} & 0.90 $\pm$ 0.00 & 0.87 $\pm$ 0.00 & 0.91 $\pm$ 0.00 & 0.90 $\pm$ 0.02 & 0.91 $\pm$ 0.00 & 0.91 $\pm$ 0.00 & 0.89 $\pm$ 0.01 & 0.88 $\pm$ 0.01 \\ \hline

\textbf{5} & 0.86 $\pm$ 0.01 & 0.85 $\pm$ 0.00 & 0.87 $\pm$ 0.00 & 0.86 $\pm$ 0.01 & 0.86 $\pm$ 0.00 & 0.85 $\pm$ 0.02 & 0.86 $\pm$ 0.01 & 0.85 $\pm$ 0.01 \\ \hline

\textbf{6} & 0.95 $\pm$ 0.01 & 0.95 $\pm$ 0.00 & 0.95 $\pm$ 0.01 & 0.95 $\pm$ 0.01 & 0.95 $\pm$ 0.00 & 0.96 $\pm$ 0.01 & \textbf{0.96} $\pm$ 0.01 & 0.95 $\pm$ 0.01 \\ \hline

\textbf{7} & 0.95 $\pm$ 0.00 & 0.95 $\pm$ 0.00 & 0.95 $\pm$ 0.01 & 0.95 $\pm$ 0.01 & 0.94 $\pm$ 0.01 & 0.94 $\pm$ 0.00 & \textbf{0.96} $\pm$ 0.01 & 0.94 $\pm$ 0.00 \\ \hline

\textbf{8} & 0.93 $\pm$ 0.00 & 0.95 $\pm$ 0.01 & 0.95 $\pm$ 0.01 & 0.95 $\pm$ 0.01 & 0.94 $\pm$ 0.00 & 0.93 $\pm$ 0.01 & 0.95 $\pm$ 0.00 & 0.94 $\pm$ 0.00 \\ \hline

\textbf{9} & 0.94 $\pm$ 0.01 & 0.94 $\pm$ 0.00 & 0.95 $\pm$ 0.01 & 0.94 $\pm$ 0.01 & 0.94 $\pm$ 0.01 & 0.94 $\pm$ 0.01 & 0.95 $\pm$ 0.01 & 0.95 $\pm$ 0.01 \\ \hline

\end{tabular}
\subcaption*{Avg: the average aggregation formula, Max: the maximum aggregation formula.}
\end{table*}

\paragraph{\textbf{Performance Analysis}} 
Regarding the embedding model, neither model showed a clear advantage. SBERT outperformed BGE in 36\% of configurations (13 out of 36), while BGE outperformed SBERT in 39\% (14 out of 36), with a tie in the remaining 25\% (9 out of 36).  
Both models achieved similar highest $F_1$-scores of 0.95 and 0.96 for BGE and SBERT, respectively. This marginal difference suggests that the choice of embedding model has a limited impact on overall performance. Because SBERT achieved a marginally higher best $F_1$-score, we selected SBERT as the preferred embedding model for requirement dependency detection. 

Regarding the similarity metric, Euclidean distance outperformed cosine similarity in 16 of 36 configurations, whereas cosine similarity outperformed Euclidean distance in 8 configurations, with a tie in 12 configurations. These results indicate that Euclidean distance provides superior similarity computation for dependency detection. Furthermore, Euclidean distance yielded higher $F_1$-scores than Cosine similarity for both BGE and SBERT.
The results also suggest that, once embeddings are generated, retrieval effectiveness is more sensitive to the choice of similarity metric than to the choice of embedding model.  

For similarity aggregation, the average aggregation formula outperformed the maximum aggregation formula in 16 of 36 configurations, while the maximum aggregation formula yielded higher $F_1$-scores in 8 configurations, with a tie in the remaining 12. 
Interestingly, the average formula provided better overall performance for both models, outperforming the maximum formula in 44.4\% of the configurations for both BGE and SBERT. 
However, with SBERT and Euclidean distance, the average formula performed substantially better than the maximum formula (67\%, with a tie in 33\% of configurations). Furthermore, both models achieved their highest $F_1$-scores with the average formula. Overall, the average formula appears to have an advantage across configurations, though this advantage is more pronounced for some combinations of embedding model and similarity metric than for others. This suggests that considering overall similarity between target and example pairs may provide more reliable guidance than relying on highly similar requirements within each pair. In particular, Average aggregation appears more stable, as it captures the broader semantic relationship between the two requirement pairs. Overall, these results indicate that the choice of similarity metric and aggregation formula had the greatest impact on the $F_1$-score, whereas the embedding model had a more modest effect. 


Finally, regarding the number of examples, the $F_1$-scores remained consistently high when the prompt included between two and four examples and between six and nine examples, although some fluctuations were observed within these ranges, as shown in \autoref{fig:FewShotConfigurations}. However, the highest $F_1$-scores were consistently achieved with six to eight examples, as reported in \autoref{tab:fewShotWithThreshold}. These findings suggest that providing a relatively large number of examples, particularly between \textbf{six and eight}, helps convey most of the useful task knowledge. The notable decline in performance when using one or five examples is primarily attributable to the severe class imbalance in the dataset. Specifically, only two instances of the \textit{details} dependency are available. Consequently, when one instance is included as an example, only a single instance remains for evaluation, causing the $F_1$-score to drop to zero if it is misclassified. Similarly, the dataset contains only six instances of the \textit{implements} dependency. When five are used as examples, only one instance remains for evaluation, making the $F_1$-score highly sensitive to a single prediction error. As noted earlier, when six or more examples are used, the \textit{Implements} dependency is excluded from the prompt because there are too few examples. These observations highlight the severity of class imbalance between dependencies and its impact on evaluation results. 

The highest $F_1$-score was achieved by three configurations, all using SBERT: (1) cosine distance with maximum aggregation and six examples, (2) Euclidean distance with average aggregation and six examples, and (3) Euclidean distance with average aggregation and seven examples. However, we selected configuration (2) because Euclidean distance with average aggregation achieved a higher $F_1$-score than the cosine distance with maximum aggregation in configuration (1), and it required fewer examples than configuration (3) (six versus seven examples per dependency type).

Thus, the optimal configuration uses \textbf{SBERT, Euclidean distance, the average similarity aggregation, 6 examples per dependency type, and a threshold-based re-annotation}. It achieved 0.96 for all evaluated metrics, including overall accuracy, precision, recall, and $F_1$-score. Performance was particularly strong for the \textit{No dependency} class, achieving an $F_1$-score of 0.98. The model also performed well across specific dependency types, achieving an $F_1$-score of 0.86 for the \textit{requires} dependency type.

\paragraph{\textbf{Robustness Analysis}}
Regarding robustness, the standard deviation remained consistently low, averaging 0.0079 across all 72 configurations. All configurations had a standard deviation of 0.02 or less, with 56.94\% at 0.01, 31.94\% with no variation across repeated executions, and 11.11\% at 0.02.  
Overall, the low standard deviation across most configurations shows strong robustness, with limited variability across repeated executions despite the inherent stochasticity of LLMs.

\subsubsection{Retrieval Augmented Generation}

Building on the optimal few-shot configurations, we integrate RAG with different numbers and chunk sizes. Table~\ref{tab:RAGwithFewShot} reports the accuracy, precision, recall, and $F_1$-score for the best-performing few-shot configuration, which uses SBERT, Euclidean distance, the average similarity aggregation, and 6 examples per dependency type. All reported results incorporate threshold-based re-annotation. 

\begin{table*}[t]
\centering
\footnotesize
\caption{Experimental Results of RAG-based Contextual Augmentation}
\label{tab:RAGwithFewShot}
\begin{tabular}{l|cc|cc|cc|c}
\hline
\multirow{2}{*}{\textbf{Metric}} & \multicolumn{2}{c|}{\textbf{2 Chunks}} & \multicolumn{2}{c|}{\textbf{6 Chunks}} & \multicolumn{2}{c|}{\textbf{10 Chunks}} & \multirow{2}{*}{\textbf{Entire Doc}} \\ \cline{2-7}
& \textbf{500c} & \textbf{1000c} & \textbf{500c} & \textbf{1000c} & \textbf{500c} & \textbf{1000c} & \\ \hline
Acc & 0.96 $\pm$ 0.00&0.96 $\pm$ 0.01& 0.97 $\pm$ 0.00&0.97 $\pm$ 0.00& 0.96 $\pm$ 0.01& 0.97 $\pm$ 0.00& 0.96 $\pm$ 0.01\\
P   & 0.96 $\pm$ 0.00&0.97 $\pm$ 0.00&  0.96 $\pm$ 0.00&0.97 $\pm$ 0.00& 0.96 $\pm$ 0.00&0.97 $\pm$ 0.00&0.96 $\pm$ 0.00\\
R   & 0.96 $\pm$ 0.00 &0.96 $\pm$ 0.01&  0.97 $\pm$ 0.00&\textbf{0.98} $\pm$ 0.00& 0.95 $\pm$ 0.01&0.97 $\pm$ 0.00&0.95 $\pm$ 0.01\\
F1  &0.96 $\pm$ 0.00&0.96 $\pm$ 0.00&  \textbf{0.97} $\pm$ 0.00 &\textbf{0.97} $\pm$ 0.00&0.96 $\pm$ 0.00& \textbf{0.97} $\pm$ 0.00&0.96 $\pm$ 0.00 \\ \hline
\end{tabular}
\smallskip
\end{table*}




As shown in \autoref{tab:RAGwithFewShot}, all configurations yielded similar results, with $F_1$-scores ranging from 0.96 to 0.97. Interestingly, precision and recall were highly balanced, suggesting that the LLM did not systematically over- or under-predict dependencies. Regarding chunk size, longer contexts (1000 characters) generally produced slightly better results than shorter ones (500 characters), with 1000-character chunks outperforming 500-character chunks when using ten chunks (0.97 vs 0.96), while also achieving higher precision (0.97 vs. 0.96) and recall (0.98 vs. 0.97) when using two and six chunks, respectively. 
This improvement likely stems from longer chunks providing more relevant information, enabling the LLM to better capture dependencies between requirements. 

The highest $F_1$-scores were achieved with 6 chunks, reaching 0.97 for both chunk sizes. Using 10 chunks also achieved an $F_1$-score of 0.97 with 1000-character chunks, suggesting that using a moderate number of chunks, namely 6 or 10, is optimal. A common pattern is that performance generally improved as the number of retrieved chunks increased from two to six or ten, before declining when the entire SRS document was included as context. These results suggest that fewer chunks may limit the diversity of retrieved context, whereas using the entire document tends to introduce noise and hinder the identification of relevant information. 

Based on these observations, although performance differences between configurations are small, we selected \textbf{six 1000-character chunks} as the optimal RAG configuration. It achieved an overall accuracy and precision of 0.97, a recall of 0.98, and an $F_1$-score of 0.97, with only three incorrect predictions out of 100 requirement pairs in the dataset. For specific dependency types, it also performed strongly, with an $F_1$-score of 0.88 for the \textit{Requires} dependency. In addition, the standard deviation remained low, with all values ranging from 0.00 to 0.01 and an average of 0.003, indicating very stable performance across repeated runs. However, given that few-shot prompting without RAG achieved an $F_1$-score of 0.96, incorporating RAG improves the $F_1$-score by only 0.01, a one-percentage-point increase. This relatively small improvement may not justify the additional computational and implementation costs associated with RAG. 

\subsubsection{\textbf{Statistical Testing}}

To assess whether the difference in performance between Few-shot and Few-shot with RAG is statistically significant, we conduct both McNemar's test \cite{McNemar_1947}and Fisher's exact test \cite{fisher1922interpretation}.
McNemar's test compares the overall predictive performance of paired approaches on the same test instances \cite{McNemar_1947}. We also apply Fisher's exact test to determine whether the observed differences in precision and recall reflect statistically significant differences in the TP/FP and TP/FN proportions \cite{fisher1922interpretation}. Throughout this paper, we use McNemar's test and Fisher's exact test to compare prompting strategies and approaches.

Considering individual runs, Fisher's exact test indicates no statistically significant difference between the Few-shot and RAG approaches in either precision or recall. All p-values are well above the conventional significance threshold of 0.05.
Similarly, McNemar’s test indicates no significant difference in prediction accuracy across any individual run. Only a few predictions differed between the two strategies: Few-shot was correct while RAG was wrong in 1–2 cases, whereas RAG was correct while Few-shot was wrong in 2–3 cases per run. 
When considering all three runs, Fisher's exact tests remain non-significant. The McNemar test is also non-significant across the three runs, despite RAG being correct in eight cases where Few-shot was incorrect and Few-shot being correct in four cases where RAG was incorrect.

Overall, these results indicate that the observed performance differences between Few-shot and RAG are not statistically significant. Together with the additional computational costs of RAG, these findings suggest that few-shot prompting alone is sufficient for dependency detection when the retrieved examples come from the same system. Based on these findings, we adopted the few-shot configuration employing \textbf{SBERT, Euclidean distance, average similarity aggregation, six examples per dependency type, and threshold-based re-annotation} for LEREDD when the few-shot examples were drawn from the same system, without RAG.

\begin{formal}
\textbf{Answer to RQ2:} Using GPT-5.4, the optimal few-shot configuration for requirement dependency detection uses SBERT, Euclidean distance, average similarity aggregation, six examples per dependency type, and threshold-based re-annotation, whereas the optimal RAG configuration uses six 1000-character chunks. Adding RAG to Few-shot does not yield statistically significant improvements over Few-shot alone, despite its additional computational cost. Few-shot prompting alone is optimal when examples originate from the same system. This configuration achieves an overall $F_1$-score of 0.96, with $F_1$-scores of 0.98 and 0.86 for \textit{No dependency} and \textit{Requires}, respectively.
\end{formal}

\subsection{\textbf{RQ3. Ablation Study of Contextual and Dynamic Example Retrieval in LEREDD}}\label{sec:RQ3Results}

To address this RQ, we conducted two analyses to evaluate: (1) the individual and combined contributions of the contextual retrieval and dynamic example retrieval stages within the LEREDD framework, and (2) the effectiveness of the semantic-based few-shot strategy adopted in the dynamic example retrieval stage compared with alternative few-shot techniques. For the first analysis, we evaluated and compared four settings: (1.1) \textit{w/o Knowledge Retrieval}, where neither stage was applied (i.e., zero-shot), (1.2) \textit{Contextual Retrieval}, where only RAG was applied, (1.3) \textit{Dynamic Examples Retrieval}, where only similarity-based few-shot was applied, and (1.4) \textit{Knowledge Retrieval}, where both RAG and similarity-based few-shot were applied. For the second analysis, we compared the similarity-based few-shot strategy with two alternatives: (2.1) \textit{Fixed Few-Shot}, where the same examples were used for all analyzed requirement pairs, and (2.2)\textit{ Random Few-Shot}, where examples were randomly selected for each analyzed requirement pair. For the few-shot and RAG strategies, we applied the optimal configurations identified in RQ2: SBERT, Euclidean distance, the average similarity aggregation formula, and six examples per dependency type for few-shot prompting, and the same few-shot configuration augmented with the retrieval of six chunks of 1000 characters for the few-shot with RAG strategy.

\subsubsection{Contribution of Contextual and Dynamic Example Retrieval within LEREDD}\autoref{tab:PromptingComparison} presents the results for both individual dependency types and overall results across the three datasets. Note that only dependency types with a sufficient number of instances (i.e., more than six instances in the dataset) were included in the comparison. This ensures that six examples are available for few-shot prompting. Consequently, we excluded the \textit{Is Similar} dependency type from D1, D2, and D3. In addition, we excluded the \textit{Details} and \textit{Conflicts} dependency types from D2 and D3, and the \textit{Implements} dependency type from D3.

\begin{table*}[t]
\centering
\scriptsize
\caption{Comparison of Different Settings Across Three Systems}
\label{tab:PromptingComparison}
\resizebox{\textwidth}{!}{
\begin{tabular}{c|c|c|ccc|c|ccc|c|ccc|c|ccc|c}
\hline
\multirow{2}{*}{\textbf{Dataset}}  & \textbf{Dependency} & \multicolumn{4}{c|}{\textbf{w/o Knowledge Retrieval}}  & \multicolumn{4}{c|}{\textbf{Contextual Retrieval}}  & \multicolumn{4}{c|}{\textbf{Dynamic Examples Retrieval}} & \multicolumn{4}{c|}{\textbf{Knowledge Retrieval}} & \multirow{2}{*}{\textbf{Supp}} \\ \cline{3-18}
 & \textbf{Type} & \textbf{Acc} & \textbf{P} & \textbf{R} & \textbf{F1} & \textbf{Acc} & \textbf{P} & \textbf{R} & \textbf{F1} & \textbf{Acc} & \textbf{P} & \textbf{R} & \textbf{F1} & \textbf{Acc} & \textbf{P} & \textbf{R} & \textbf{F1} \\ \hline

\multirow{5}{*}{D1}
& No Dependency  & \multirow{6}{*}{\makecell{0.75}} &0.87&0.86&0.86& \multirow{6}{*}{\makecell{0.79}} & 0.88 & 0.89& 0.88 & \multirow{6}{*}{\makecell{0.91}} &0.93&0.97&0.95& \multirow{6}{*}{\makecell{0.91}} &0.93&0.97&0.95& 229\\
& Requires     & &0.34&0.40&0.37&  & 0.51& 0.53& 0.52&  &0.91&0.73&0.81& &0.90&0.73&0.81& 40\\
& Implements   & &0.60&0.70&0.64&  & 0.68& 0.63& 0.65&  &0.73&0.80&0.76& &0.75&0.87&0.80& 18\\
& Details      & &0.62&0.41&0.49&  & 0.45& 0.33&0.38&  &1.00&0.79&0.89& &0.94&0.74&0.83& 13\\
& Conflicts    & &0.53&0.30&0.38&  & 0.55& 0.47& 0.50& &0.86&0.60&0.71& &0.94&0.50&0.65& 10\\
\cline{2-2} \cline{4-6} \cline{8-10} \cline{12-14} \cline{16-19}
& \textit{Overall} & &0.76&0.75&0.76&  & 0.79& 0.79& 0.79 & &0.91&0.91&\textbf{0.91} & & 0.92&0.91&0.91 & 310\\
\hline

\multirow{4}{*}{D2}
& No Dependency      &  \multirow{4}{*}{\makecell{0.85}} &0.96&0.91&0.93& \multirow{4}{*}{\makecell{0.87}} & 0.99 & 0.86& 0.93 &\multirow{4}{*}{\makecell{0.95}} &0.96&0.98&0.97&\multirow{4}{*}{\makecell{0.93}} &0.96&0.97&0.97& 170\\
& Requires     & &0.46&0.61&0.52&  & 0.52& 0.76& 0.62& &0.88&0.71&0.78& &0.85&0.67&0.75& 17\\
& Implements   & &0.34&0.37&0.35&  & 0.33& 0.80& 0.47& &0.78&0.80&0.79& &0.68&0.70&0.69& 10\\
\cline{2-2} \cline{4-6} \cline{8-10} \cline{12-14} \cline{16-19}
& \textit{Overall} & &0.88&0.85&0.87&  & 0.91& 0.85&  0.88& &0.94&0.94&\textbf{0.94}& &0.93&0.93&0.93& 197\\
\hline

\multirow{3}{*}{D3}
& No Dependency      &  \multirow{3}{*}{\makecell{0.68}} &0.98&0.66&0.79& \multirow{3}{*}{\makecell{0.81}} & 0.99& 0.76& 0.86& \multirow{3}{*}{\makecell{0.86}} &0.95&0.86&0.90& \multirow{3}{*}{\makecell{0.85}}&0.96&0.84&0.89& 144 \\
& Requires     & &0.45&0.76&0.57&  & 0.56& 0.98& 0.71& &0.65&0.87&0.75& &0.63&0.89&0.74& 45\\
\cline{2-2} \cline{4-6} \cline{8-10} \cline{12-14} \cline{16-19}
& \textit{Overall} & &0.85&0.68&0.73&  &0.89& 0.81& 0.82& &0.87&0.86&\textbf{0.87}& &0.88&0.85&0.86& 189\\
\hline

\end{tabular}}
\end{table*}

\paragraph{\textbf{Performance Analysis}} 
As \autoref{tab:PromptingComparison} shows, \textit{w/o Knowledge Retrieval setting} yields the lowest accuracy and $F_1$-scores across all three datasets, highlighting the LLM's need for contextual information for the task. Its accuracy is lower than that of the other settings, with values of 0.75, 0.85, and 0.68 for D1, D2, and D3, respectively, compared to 0.79, 0.87, and 0.81 for \textit{Contextual Retrieval}, 0.91, 0.95, and 0.86 for \textit{Dynamic Examples Retrieval}, and 0.91, 0.93, and 0.85 for \textit{Knowledge Retrieval}. Similarly, its overall $F_1$-scores are lower than the other settings, ranging from 0.73 for D3 to 0.87 for D2. Although the overall $F_1$-scores appear high, the performance on specific dependency types is considerably low. This is not reflected in the overall $F_1$-score because it averages across all classes, and specific dependency types occur much less frequently than the \textit{No dependency} class. The $F_1$-score varies considerably across specific dependency types, yielding scores as low as 0.35 for \textit{implements} in D2 and 0.37 for \textit{requires} in D1. Across the three datasets, the $F_1$-scores for specific dependency types average 0.47. These results confirm the observations made in RQ1, indicating that zero-shot LLMs particularly struggle to identify specific dependency types. Furthermore, recall is consistently higher than precision for specific dependency types, whereas precision is higher than recall for \textit{No dependency} across all three datasets. This suggests that the zero-shot model tends to over-predict specific dependency types, resulting in a relatively high number of false positives.
Notably, the $F_1$-scores for \textit{No dependency} are high, reaching 0.86, 0.93, and 0.79 for D1, D2, and D3, respectively. These scores are only 6\%, 4\%, and 11\% lower than those achieved by the best-performing setting. 

\textit{Contextual Retrieval} performs better than the \textit{w/o Knowledge Retrieval} setting, but considerably worse than the other two settings. Specifically, it achieves $F_1$-scores slightly higher than \textit{w/o Knowledge Retrieval} for D1 and D2 (0.79 and 0.88 compared to 0.76 and 0.87, respectively), while the difference is more pronounced for D3 (0.82 compared to 0.73). This difference may be due to D3 having fewer dependency types than D1 and D2. 
The $F_1$-scores for specific dependency types were higher in almost all cases, averaging 0.55 compared to 0.47 for \textit{w/o Knowledge Retrieval}. Nevertheless, across the three datasets, \textit{Contextual Retrieval} achieves an average $F_1$ of 0.83 compared to 0.79 for \textit{w/o Knowledge Retrieval}, indicating relatively similar overall performance. These results suggest that incorporating domain context through RAG provides only limited improvements over zero-shot, as domain context alone does not provide sufficient information to effectively guide the LLM in detecting requirement dependencies. 

The performance differences between \textit{Contextual Retrieval} and the two few-shot settings are considerably more pronounced. The \textit{Contextual Retrieval} setting achieves an average $F_1$-score of 0.83, compared to 0.91 for \textit{Dynamic Examples Retrieval} and 0.90 for \textit{Knowledge Retrieval}. While the $F_1$-scores for \textit{No dependency} are higher in the latter two settings, consistently reaching at least 0.90, the difference is more pronounced for specific dependency types, for which the average $F_1$-scores are 0.78 and 0.75 for \textit{Dynamic Examples Retrieval} and \textit{Knowledge Retrieval}, respectively, compared to 0.55 for \textit{Contextual Retrieval}. These results highlight that examples provide more effective guidance than domain context alone within LEREDD for requirement dependency detection. In particular, concrete examples can be particularly useful for accurately identifying specific dependency types.

When comparing \textit{Dynamic Examples Retrieval} with \textit{Knowledge Retrieval}, the results are very similar, with \textit{Dynamic Examples Retrieval} achieving a slightly higher $F_1$-score in two of the three datasets. Specifically, the average $F_1$-scores reached 0.91, 0.94, and 0.87 for \textit{Dynamic Examples Retrieval}, and 0.91, 0.93, and 0.86 for \textit{Knowledge Retrieval}, for D1, D2, and D3, respectively. While the $F_1$-score was the same for D1, \textit{Dynamic Examples Retrieval} achieved a marginally higher score for D2 and D3 (+1\%). Furthermore, the $F_1$-score for \textit{No dependency} remains generally consistent across the two settings, with the main difference being in specific dependency types. \textit{Dynamic Examples Retrieval} achieved higher $F_1$-scores for most specific dependency types. For example, in D2, the $F_1$-scores for \textit{Requires} and \textit{Implements} increased from 0.69 to 0.79 and from 0.75 to 0.78, respectively.
The $F_1$-scores of specific dependencies for the \textit{Dynamic Examples Retrieval} setting ranged between 0.71 and 0.89 across the three datasets. This suggests that adding RAG primarily affects the classification of specific dependency types rather than whether a dependency exists.

Together, these results and the efficiency gains from not using RAG suggest that \textit{Dynamic Examples Retrieval} alone offers a better trade-off between performance and efficiency. One possible explanation is that the additional context introduced by RAG provides limited benefit when the prompt already includes six examples from the same system per dependency type. The additional retrieved context may instead introduce unnecessary information and noise that could negatively affect the LLM's classification.


\paragraph{\textbf{Confidence Score Analysis}}
In addition to the evaluation metrics, we compared the average confidence scores across the four settings. \autoref{tab:confidenceScore} presents the results for confidence scores of 5 and 4, as these account for the majority of the reported scores. Consistent with the results presented in \autoref{tab:PromptingComparison}, the \textit{Dynamic Examples Retrieval} setting yielded the highest percentage of confidence scores equal to 5, closely followed by the \textit{Knowledge Retrieval} setting, followed by \textit{Contextual Retrieval}, while the \textit{w/o Knowledge Retrieval} setting yielded the lowest percentage. This percentage was similar for the two few-shot settings, exceeding that of \textit{Knowledge Retrieval} by only 2\%, 2\%, and 1\% for D1, D2, and D3, respectively. In contrast, the difference between the \textit{Dynamic Examples Retrieval} and \textit{w/o Knowledge Retrieval} settings was significant: the percentage of confidence scores equal to 5 decreased from 97\% to 59\% for D1, from 98\% to 63\% for D2, and from 98\% to 35\% for D3. While almost all confidence scores reported by the \textit{Dynamic Examples Retrieval} and \textit{Knowledge Retrieval} settings were 5, the \textit{w/o Knowledge Retrieval} setting produced a more balanced distribution between scores of 5 and 4, with the percentage of confidence scores of 4 even exceeding the percentage of score 5 for D3. A similar but less pronounced pattern was observed for \textit{Contextual Retrieval}, where the percentage of confidence scores of 5 decreased from an average of 98\% for \textit{Dynamic Examples Retrieval} to 67\%, a difference of more than 30 percentage points.  These results suggest that zero-shot and RAG-only augmented models are much less confident in their predictions than the other two strategies, likely because they lack task-specific guidance, which increases uncertainty when determining the dependency type. In contrast, the additional guidance and contextual information from examples give the model more relevant information for making its predictions, thereby increasing its reported confidence.

\begin{table}[t]
\centering
\scriptsize
\caption{Comparison of Confidence Scores of Different Settings Across Three Datasets}
\label{tab:confidenceScore}
\resizebox{\columnwidth}{!}{
\begin{tabular}{p{0.7cm}|p{1.1cm}|p{1.2cm}|p{1.05cm}|p{1cm}|p{1cm}}
\hline

\textbf{Dataset} & \textbf{Confidence score} & \textbf{\textit{w/o Knowledge Retrieval}} & \textbf{\textit{Contextual Retrieval}}& \textbf{\textit{Dynamic Example Retrieval}} & \textbf{\textit{Knowledge Retrieval}} \\ \hline
\multirow{2}{*}{D1} & 5 &  59\%& 71\% & \textbf{97\%} & 95\% \\ 
 & 4 & 36\%& 28\% & 3\% & 4\% \\ \hline 
\multirow{2}{*}{D2} & 5 & 63\% & 66\% & \textbf{98\%} & 96\% \\ 
 & 4 & 36\% & 33\% & 2\% & 4\% \\ \hline 
 \multirow{2}{*}{D3} & 5 & 35\% & 64\% & \textbf{98\%} & 97\% \\ 
 & 4 & 58\% & 36\% & 2\% & 2\% \\ \hline
\end{tabular}}
\end{table}

\paragraph{\textbf{Statistical Testing}}

Across the three datasets, both \textit{Dynamic Examples Retrieval} and \textit{Knowledge Retrieval} significantly outperform \textit{w/o Knowledge Retrieval} and \textit{Contextual Retrieval}. For D1, Fisher's exact tests indicate that \textit{Dynamic Examples Retrieval} significantly outperforms \textit{w/o Knowledge Retrieval} and \textit{Contextual Retrieval}, while the McNemar test also indicates a significant difference in overall prediction accuracy. When comparing \textit{Knowledge Retrieval} with \textit{w/o Knowledge Retrieval} and \textit{Contextual Retrieval}, the precision difference was also significant. The McNemar test further confirms a significant difference in overall accuracy.
For D2, Fisher's exact tests indicate that both \textit{Dynamic Examples Retrieval} and \textit{Knowledge Retrieval} significantly outperform \textit{w/o Knowledge Retrieval} in precision but not in recall. The McNemar test confirms a significant difference in overall accuracy when compared to \textit{Dynamic Examples Retrieval} and \textit{Knowledge Retrieval}. Finally, for D3, the McNemar test indicates a significant difference in overall accuracy for both \textit{Dynamic Examples Retrieval} and \textit{Knowledge Retrieval} compared with \textit{w/o Knowledge Retrieval}. Overall, these results demonstrate a clear performance advantage of both Few-shot approaches over Zero-shot.

In contrast, \textit{Dynamic Examples Retrieval} and \textit{Knowledge Retrieval} do not differ significantly across the three datasets. The Fisher's exact tests were not significant for precision and recall. The McNemar test further confirmed these results. Thus, adding RAG does not seem to improve Few-shot performance across the three datasets.
These findings confirm the results obtained in RQ2.

\paragraph{\textbf{Robustness Analysis}}
Regarding robustness, the standard deviation was generally low, averaging 0.016 across all settings and datasets. However, variability differed across settings. \textit{w/o Knowledge Retrieval} and \textit{Contextual Retrieval} exhibited the highest average deviation (0.021), followed by \textit{Knowledge Retrieval} (0.013) and \textit{Dynamic Example Retrieval} (0.009). These results suggest that incorporating examples improves stability compared with zero-shot and RAG-only augmented models, while adding RAG context to few-shot introduces slightly greater variability than few-shot prompting alone.
Furthermore, only a few deviations were 0.05 or higher, and we observed these for \textit{w/o Knowledge Retrieval}, \textit{Contextual Retrieval}, and \textit{Knowledge Retrieval}, particularly for specific dependency types. For example, the recall for the \textit{Details} dependency in D1 exhibited a standard deviation of 0.10 under the \textit{w/o Knowledge Retrieval} setting. These higher deviations likely stem from the limited number of instances for some dependency types, where small variations in predictions across runs can substantially affect the evaluated metrics. Overall, \textit{Dynamic Example Retrieval} showed the most stable performance across repeated executions, while incorporating RAG for \textit{Contextual Retrieval} led to slightly higher variability. Nevertheless, the relatively low deviations across the configurations indicate generally stable performance despite the inherent stochasticity of LLMs.


\subsubsection{Effectiveness of Semantic-Based Few-Shot}

\autoref{tab:FewShotPromptingComparison} presents the results of similarity-based few-shot prompting compared with alternative few-shot techniques: (1) fixed few-shot prompting, where the same examples are used for all pairs, and (2) random few-shot prompting, where the examples are randomly selected for each pair.

\begin{table*}[t]
\centering
\scriptsize
\caption{Comparison of Different Few-shot Strategies Across Three Systems}
\label{tab:FewShotPromptingComparison}
\resizebox{\textwidth}{!}{

\begin{tabular}{c|c|c|ccc|c|ccc|c|ccc|c}
\hline
\multirow{2}{*}{\textbf{Dataset}}  & \textbf{Dependency} & \multicolumn{4}{c|}{\textbf{Fixed Few-shot}}  & \multicolumn{4}{c|}{\textbf{Random Few-shot}}  & \multicolumn{4}{c|}{\textbf{Similarity-based Few-shot}} & \multirow{2}{*}{\textbf{Supp}} \\ \cline{3-14}
 & \textbf{Type} & \textbf{Acc} & \textbf{P} & \textbf{R} & \textbf{F1} & \textbf{Acc} & \textbf{P} & \textbf{R} & \textbf{F1} & \textbf{Acc} & \textbf{P} & \textbf{R} & \textbf{F1}  \\ \hline

\multirow{5}{*}{D1}
& No Dependency  & \multirow{6}{*}{\makecell{0.81  \\$\pm$0.00}} &0.96$\pm$0.00&0.82$\pm$0.00&0.88$\pm$0.00& \multirow{6}{*}{\makecell{0.77 \\$\pm$0.02}} &0.95$\pm$0.01&0.78$\pm$0.01&0.86$\pm$0.01& \multirow{6}{*}{\makecell{0.91 \\$\pm$0.00}} &0.93$\pm$0.00&0.97$\pm$0.00&0.95$\pm$0.00& 229\\
& Requires       & &0.60$\pm$0.01&0.77$\pm$0.01&0.67$\pm$0.01& &0.52$\pm$0.02&0.68$\pm$0.04&0.59$\pm$0.03& &0.91$\pm$0.00&0.73$\pm$0.01&0.81$\pm$0.01& 40\\
& Implements     & &0.56$\pm$0.03&0.78$\pm$0.05&0.65$\pm$0.04& &0.44$\pm$0.02&0.80$\pm$0.03&0.57$\pm$0.03& &0.73$\pm$0.02&0.80$\pm$0.03&0.76$\pm$0.02& 18\\
& Details        & &0.57$\pm$0.02&0.87$\pm$0.04&0.69$\pm$0.03& &0.51$\pm$0.05&0.87$\pm$0.04&0.64$\pm$0.05& &1.00$\pm$0.00&0.79$\pm$0.04&0.89$\pm$0.02 & 13\\
& Conflicts      & &0.35$\pm$0.02&0.60$\pm$0.00&0.44$\pm$0.02& &0.44$\pm$0.07&0.63$\pm$0.05&0.52$\pm$0.07& &0.86$\pm$0.00&0.60$\pm$0.00&0.71$\pm$0.00  & 10\\
\cline{2-2} \cline{4-6} \cline{8-10} \cline{12-15}
& \textit{Overall} &  &0.85$\pm$0.00&0.81$\pm$0.00&0.82$\pm$0.00& &0.83$\pm$0.01&0.77$\pm$0.02&0.79$\pm$0.02& &0.91$\pm$0.00&0.91$\pm$0.00 &\textbf{0.91}$\pm$0.00 & 310\\
\hline

\multirow{4}{*}{D2}
& No Dependency  & \multirow{4}{*}{\makecell{0.86 \\$\pm$0.01 }} &0.99$\pm$0.00&0.86$\pm$0.01&0.92$\pm$0.00& \multirow{4}{*}{\makecell{0.87 \\$\pm$0.01}} &0.98$\pm$0.00&0.87$\pm$0.01&0.93$\pm$0.01& \multirow{4}{*}{\makecell{0.95 \\$\pm$0.00}} &0.96$\pm$0.00&0.98$\pm$0.00&0.97$\pm$0.00& 170\\
& Requires       & &0.57$\pm$0.04&0.90$\pm$0.03&0.70$\pm$0.04& &0.61$\pm$0.04&0.80$\pm$0.07&0.69$\pm$0.02& &0.88$\pm$0.03&0.71$\pm$0.00&0.78$\pm$0.01& 17\\
& Implements     & &0.38$\pm$0.03&0.87$\pm$0.05&0.53$\pm$0.03& &0.38$\pm$0.08&0.87$\pm$0.05&0.52$\pm$0.08& &0.78$\pm$0.03&0.80$\pm$0.00&0.79$\pm$0.02& 10\\
\cline{2-2} \cline{4-6} \cline{8-10} \cline{12-15}
& \textit{Overall} & &0.93$\pm$0.00&0.86$\pm$0.01&0.88$\pm$0.01& &0.92$\pm$0.00&0.87$\pm$0.01&0.88$\pm$0.01& &0.94$\pm$0.00&0.94$\pm$0.00&\textbf{0.94}$\pm$0.00& 197\\
\hline

\multirow{3}{*}{D3}
& No Dependency  & \multirow{3}{*}{\makecell{0.79 \\$\pm$0.00}} &0.94$\pm$0.01&0.77$\pm$0.01&0.85$\pm$0.00& \multirow{3}{*}{\makecell{0.77 \\$\pm$0.01}} &0.98$\pm$0.00&0.72$\pm$0.01&0.83$\pm$0.01& \multirow{3}{*}{\makecell{0.86 \\$\pm$0.01}} &0.95$\pm$0.01&0.86$\pm$0.00&0.90$\pm$0.01& 144\\
& Requires       & &0.54$\pm$0.01&0.84$\pm$0.02&0.66$\pm$0.01& &0.51$\pm$0.01&0.96$\pm$0.00&0.67$\pm$0.01& &0.65$\pm$0.01&0.87$\pm$0.03&0.75$\pm$0.02& 45\\
\cline{2-2} \cline{4-6} \cline{8-10} \cline{12-15}
& \textit{Overall} & &0.84$\pm$0.01&0.79$\pm$0.00&0.80$\pm$0.00& &0.87$\pm$0.00&0.77$\pm$0.01&0.79$\pm$0.01& &0.87$\pm$0.01&0.86$\pm$0.01&\textbf{0.87}$\pm$0.01& 189\\
\hline

\end{tabular}}
\end{table*}

\paragraph{\textbf{Performance Analysis}} 

When comparing fixed and random few-shot, performance is comparable across datasets, with slightly higher scores for fixed few-shot. Specifically, fixed few-shot achieved $F_1$-scores of 0.82, 0.88, and 0.80, compared to 0.79, 0.88, and 0.79 for random few-shot on D1, D2, and D3, respectively. The slightly higher performance of fixed few-shot is attributed to marginally higher $F_1$-scores for both No-dependency and specific dependency types. In fact, the average $F_1$-score for specific dependency types across the three datasets is 0.62 for fixed few-shot, compared to 0.60 for random few-shot. These results suggest that using the same examples across all pairs or randomly selecting examples for each pair leads to relatively similar performance overall.

The difference is considerably more pronounced when comparing these two techniques with similarity-based few-shot. The latter outperforms both fixed and random few-shot across datasets and dependency types, achieving an average $F_1$-score of 0.91, compared to 0.83 for fixed few-shot and 0.82 for random few-shot. Although this improvement is also reflected in the classification of \textit{No dependency}, for which the $F_1$-score consistently exceeds 0.90, it is more pronounced for specific dependency types. The average $F_1$-score for these types reaches 0.78 with similarity-based few-shot, compared to 0.62 for fixed few-shot and 0.60 for random few-shot. 
These results suggest that selecting semantically similar examples provides more effective guidance to the LLM for detecting requirement dependencies. This selection appears to improve both identifying specific dependency types and classifying whether a dependency exists. This improvement may be attributed to semantically similar examples being more representative of the analyzed requirements, making it easier for the LLM to distinguish between dependency types. Furthermore, these examples could provide valuable domain information about the same or similar entities mentioned in the requirements, helping the LLM better understand their content and, consequently, identify dependencies more accurately.

\paragraph{\textbf{Confidence Score Analysis}}

\autoref{tab:confidenceScoreFewShot} presents the distribution of confidence scores 5 and 4 across the three few-shot selection strategies and datasets. As the table shows, fixed and random few-shot yielded comparable distributions for both scores. Specifically, fixed few-shot yielded averages of 68\% and 29\% for scores 5 and 4, respectively, compared with 69\% and 29\% for random few-shot. These results suggest that randomly selected examples provide comparable confidence levels regardless of whether the same examples are used across all requirement pairs or different examples are selected for each pair. These moderate confidence levels may be attributed to the limited domain-specific guidance provided by randomly selected examples. In contrast, similarity-based few-shot yielded substantially higher confidence scores. It achieved an average of 98\% for score 5, representing an increase of approximately 30 percentage points compared with the other two strategies. This higher confidence may be because semantically similar examples better represent the requirements being analyzed, giving the LLM more specific guidance for dependency detection and increasing its reported confidence.

\begin{table}[t]
\centering
\footnotesize
\caption{Comparison of Confidence Scores of Different Few-shot Prompting Strategies Across Three Datasets}
\label{tab:confidenceScoreFewShot}
\resizebox{\columnwidth}{!}{
\begin{tabular}{c|c|c|c|c}
\hline

\textbf{Dataset} & \textbf{Confidence score} & \textbf{Fixed} & \textbf{Random}& \textbf{Similarity-based} \\ \hline
\multirow{2}{*}{D1} & 5 &  67\%& 67\% & \textbf{97\%} \\ 
 & 4 & 31\%& 29\% & 3\%  \\ \hline 
\multirow{2}{*}{D2} & 5 & 75\% & 71\% & \textbf{98\%}  \\ 
 & 4 & 24\% & 27\% & 2\% \\ \hline 
 \multirow{2}{*}{D3} & 5 & 63\% & 68\% & \textbf{98\%}  \\ 
 & 4 & 34\% & 30\% & 2\% \\ \hline
\end{tabular}}
\end{table}

\paragraph{\textbf{Statistical testing}}

Across datasets, fixed few-shot and random few-shot generally do not exhibit significant differences. Specifically, for D2 and D3, Fisher's exact tests and the McNemar test indicate no significant differences in precision, recall, or overall prediction accuracy. For D1, Fisher's exact test indicates that fixed few-shot achieves significantly higher prediction accuracy than random few-shot, while precision and recall show no significant differences. Overall, these results suggest that fixed and random few-shot strategies yield statistically comparable performance.

Similarity-based few-shot achieves significantly better performance than the other two strategies. Across all three datasets, Fisher's exact tests indicate that similarity-based few-shot achieves significantly higher precision than both fixed and random few-shot, whereas recall differences are not statistically significant. Furthermore, the McNemar test indicates significantly higher prediction accuracy for similarity-based few-shot compared with both alternatives. These results suggest that similarity-based example selection improves performance, particularly in terms of precision and overall prediction accuracy.

\paragraph{\textbf{Robustness Analysis}}

As \autoref{tab:FewShotPromptingComparison} shows, the standard deviations are generally low, averaging 0.016 across all few-shot strategies and datasets. However, variability differs across strategies. Random few-shot has the highest average standard deviation (0.025), followed by fixed few-shot (0.015) and similarity-based few-shot (0.009). Notably, none of the standard deviations for similarity-based few-shot reach 0.05, whereas two do for fixed few-shot and nine do for random few-shot. The higher variability observed with random few-shot may result from the examples being selected differently across runs, potentially leading to variations in the LLM's predictions. In contrast, fixed few-shot uses the same examples across all requirement pairs and runs, thereby reducing variability. Finally, although similarity-based few-shot also retrieves the same examples for a given requirement pair across runs, it additionally selects semantically similar examples. These examples can provide more specific guidance for dependency detection, contributing to more stable results. Overall, although standard deviations are generally low across all strategies, similarity-based few-shot consistently exhibits the lowest variability, further highlighting the benefits of similarity-based example selection.

\begin{formal}
\textbf{Answer to RQ3:} The \textit{Dynamic Examples Retrieval} stage makes the most substantial contribution to LEREDD's dependency detection performance for both \textit{No dependency} and specific dependency types. Across datasets, it achieves an overall $F_1$-score of 0.91, with average $F_1$-scores of 0.94 and 0.79 for \textit{No dependency} and specific dependency types, yielding relative gains of 9.3\% and 66\%, respectively, compared to zero-shot. The differences in precision, recall, and overall accuracy were statistically significant. Furthermore, adding \textit{Contextual Retrieval} to \textit{Dynamic Examples Retrieval} does not improve performance. This confirms that RAG provides no additional performance advantage over Few-shot alone when examples come from the same system. Moreover, semantic-based example selection in \textit{Dynamic Examples Retrieval} yields statistically significant improvements over fixed and random few-shot selection, highlighting the importance of dynamically retrieving semantically relevant examples for LEREDD.
\end{formal}

\subsection{\textbf{RQ4. Comparison of LEREDD with Baselines for Intra-Dataset Dependency Detection}}\label{sec:RQ4Results}

To assess the performance of LEREDD against SOTA approaches, we compare it to two baselines for requirement dependency detection: (1) TF-IDF \& LSA as a retrieval-based approach, and (2) fine-tuned BERT as an ML-based approach. In this section, we evaluate these approaches for intra-dataset dependency detection, where the instances used to fine-tune BERT and the examples provided to LEREDD originate from the same system as the test set. 
Furthermore, because the implemented TF-IDF \& LSA approach is designed only to recommend \textit{requires} dependencies, we evaluated only this dependency type as a baseline.

\begin{table*}[t]
\centering
\footnotesize
\caption{Experimental Results of LEREDD Against Baselines for Intra-Dataset Evaluation across Three Systems}
\label{tab:intraDatasetResults}
\resizebox{0.9\textwidth}{!}{
\begin{tabular}{c|c|c|ccc|c|ccc|c|ccc|c}
\hline
\multirow{2}{*}{\textbf{System}} & \textbf{Dependency} & \multicolumn{4}{c|}{\textbf{TF-IDF \& LSA}} & \multicolumn{4}{c|}{\textbf{fine-tuned BERT}} & \multicolumn{4}{c|}{\textbf{LEREDD}}& \multirow{2}{*}{\textbf{Supp}} \\ \cline{3-14} 
 & \textbf{Type} & \textbf{Acc} & \textbf{P} & \textbf{R} & \textbf{F1} & \textbf{Acc} & \textbf{P} & \textbf{R} & \textbf{F1} & \textbf{Acc} & \textbf{P} & \textbf{R} & \textbf{F1}  \\ \hline
\multirow{6}{*}{D1} & No Dependency& \multirow{6}{*}{0.68} & 0.88 & 0.73 & 0.80 & \multirow{6}{*}{0.87} & 0.93 & 0.93 & \textbf{0.93} & \multirow{6}{*}{\textbf{0.88} $\pm$   0.01} &0.92 $\pm$    0.01&0.94 $\pm$    0.02&\textbf{0.93} $\pm$    0.01& 48\\
 & Requires & &0.19  & 0.38 & 0.25 & & 0.86 & 0.75 & \textbf{0.80} & &0.83 $\pm$    0.02&0.62  $\pm$   0.10&0.71  $\pm$  0.08 & 8\\
 & Implements &  & N/A & N/A & N/A &  & 0.50  & 0.67 & 0.57 & &0.70  $\pm$   0.07&1.00  $\pm$   0.00&\textbf{0.82}  $\pm$   0.05& 3\\ 
 & Details &  & N/A & N/A & N/A &  & 0.67 & 0.67 & 0.67 & &1.00  $\pm$   0.00&0.67  $\pm$   0.00&\textbf{0.80}  $\pm$   0.00& 3\\ 
  & Conflicts &  &  N/A & N/A & N/A &  & 0.50 & 0.50 & \textbf{0.50} & &0.44  $\pm$   0.08&0.50  $\pm$   0.00&0.47  $\pm$   0.05& 2\\ \cline{2-2} \cline{4-6} \cline{8-10} \cline{12-15} 
 & \textit{Overall} &  & 0.78 & 0.68 & 0.72 &  & 0.87 & 0.87 & 0.87 &   &\textbf{0.88}  $\pm$   0.01&\textbf{0.88}  $\pm$   0.00&\textbf{0.88}  $\pm$   0.01& 64\\ \hline
 
\multirow{3}{*}{D2} & No Dependency& \multirow{3}{*}{0.74} & 0.96 & 0.74 &  0.83 & \multirow{3}{*}{0.85} & 0.91 & 0.94 & 0.93 & \multirow{3}{*}{\textbf{0.88}  $\pm$  0.00} &0.93  $\pm$   0.01&0.94  $\pm$   0.00&\textbf{0.94}  $\pm$   0.01& 34\\
 & Requires & & 0.25 & 0.75 & 0.38 & & 0.50 & 0.50 & 0.50 & &0.78  $\pm$   0.16&0.50  $\pm$   0.00&\textbf{0.60}  $\pm$   0.04& 4\\
 & Implements & & N/A & N/A & N/A & & 0.00 & 0.00 & 0.00 & &0.33  $\pm$   0.00&0.50  $\pm$   0.00&\textbf{0.40}  $\pm$   0.00& 2\\  \cline{2-2} \cline{4-6} \cline{8-10} \cline{12-15} 
 & \textit{Overall} &  &  0.88 &  0.74 & 0.78 &  & 0.83 & 0.85 & 0.84 &  &\textbf{0.88}  $\pm$   0.00&\textbf{0.88}  $\pm$   0.00&\textbf{0.88}  $\pm$   0.00& 40\\ \hline
 
\multirow{3}{*}{D3} & No Dependency& \multirow{3}{*}{0.46} & 0.76 & 0.43 & 0.55 & \multirow{3}{*}{0.67} & 0.95 & 0.60 & 0.73 & \multirow{3}{*}{\textbf{0.87} $\pm$   0.00} &0.93  $\pm$   0.00&0.90  $\pm$   0.00&\textbf{0.92}  $\pm$   0.00& 30\\
 & Requires & & 0.23 & 0.56 & 0.32 & & 0.40 & 0.89 & 0.55 & &0.70  $\pm$   0.00&0.78  $\pm$   0.00&\textbf{0.74}  $\pm$   0.00& 9\\ \cline{2-2} \cline{4-6} \cline{8-10} \cline{12-15} 
 & \textit{Overall} &  & 0.63 & 0.46 & 0.49 &  & 0.82 & 0.67 & 0.69 &  &\textbf{0.88}  $\pm$   0.00&\textbf{0.87}  $\pm$   0.00&\textbf{0.87}  $\pm$   0.00& 39\\  \hline
\end{tabular}}
\centering

\footnotesize
Note: N/A indicates dependency types not supported by TF-IDF \& LSA.

\end{table*}

\paragraph{\textbf{Overall Performance}} 
Table~\ref{tab:intraDatasetResults} reports the intra-dataset comparison between LEREDD and the two baselines. The highest results are highlighted in bold, indicating that LEREDD consistently achieves the highest accuracy and $F_1$-scores across all dependency types compared with the two baselines. LEREDD achieves the highest accuracy and $F_1$-scores on all three systems, reaching 0.88 on D1, 0.88 on D2, and 0.87 on D3. As expected, these results are lower than those in the previous RQs because the 80\%/20\% train-test split substantially reduces the number of instances for each dependency type, with counts ranging from 2 to 9 per type. Consequently, the performance metrics become considerably more sensitive to individual misclassifications. Nevertheless, LEREDD still consistently outperforms the strongest baseline (BERT) on each dataset. We observe relative $F_1$-score improvements of 1.15\% on D1 (from 0.87 to 0.88), 4.76\% on D2 (from 0.84 to 0.88), and 26.09\% on D3 (from 0.69 to 0.87), respectively. Across the three datasets, LEREDD achieved the highest average $F_1$-score (0.88), followed by fine-tuned BERT (0.80) and TF-IDF \& LSA (0.66). 

\paragraph{\textbf{Performance on Specific Dependency Types}}

Regarding \textit{No dependency}, LEREDD achieved very high $F_1$-scores ranging from 0.92 to 0.94 across all datasets. Fine-tuned BERT achieved comparable performance on D1 and D2 but dropped substantially on D3, with the $F_1$-score falling to 0.73. Similarly, TF-IDF \& LSA performed well on D1 and D2, reaching $F_1$-scores of 0.80 and 0.83, respectively, but it also exhibited a notable decline on D3, with an $F_1$-score of 0.55. Overall, although all three approaches perform well in identifying \textit{No dependency} instances, LEREDD consistently provides the highest performance, with an average $F_1$-score of 0.93 across the three datasets.

Despite the limited number of instances for several dependency types, LEREDD consistently outperformed the two baselines. The $F_1$-scores for most dependency types were above 0.70, with only two cases below 0.50, both for dependency types represented by only two instances. Although fine-tuned BERT achieved the highest $F_1$-score for two dependency types, its performance was less consistent than LEREDD, often yielding $F_1$-scores around 0.50 and dropping to 0.00 for the \textit{Implements} dependency in D2. These results highlight the limitations of ML-based approaches in handling severe class imbalance. Similarly, TF-IDF \& LSA achieved substantially lower performance for the \textit{requires} dependency, with $F_1$-scores ranging from 0.25 to 0.32. These results highlight LEREDD's superior performance and greater robustness to the inherent class imbalance among dependency types. For instance, for the \textit{Requires} dependency, LEREDD improves the $F_1$-score from 0.32 and 0.55 to 0.74 on D3, yielding relative gains of 131.25\% and 34.55\% over TF-IDF \& LSA and fine-tuned BERT, respectively. Across systems, it yields average relative gains of 124\% and 14.43\% in $F_1$-scores for the \textit{Requires} dependency over the baselines and provides a better balance between precision and recall. For the \textit{implements} dependency in D1, which contains only a few instances, LEREDD attains an $F_1$-score of 0.82, representing a 43.86\% improvement over fine-tuned BERT. Furthermore, the baselines exhibit noticeable instability across datasets. For example, the accuracy of TF-IDF \& LSA and BERT drops from 0.68 and 0.87 on D1 to 0.46 and 0.67 in D3, whereas LEREDD remains consistently above 0.87.

\paragraph{\textbf{Statistical Testing}}

When comparing LEREDD with TF-IDF \& LSA, LEREDD demonstrates statistically significant differences. Overall, Fisher's tests show significant differences in both precision and recall. More specifically, these results indicate that LEREDD significantly outperforms TF-IDF \& LSA, particularly in D1 and D3. For D2, McNemar’s test shows a significant difference in prediction accuracy. 

Compared with BERT, LEREDD generally performs comparably across two of the three datasets, although significant differences emerge on specific metrics. In D1, neither Fisher's exact test nor McNemar's test indicates a statistically significant difference, with all $p$-values above 0.05. In D2, Fisher's exact test shows no significant difference in precision, but a significant difference in recall. However, McNemar's test remains non-significant, indicating no significant difference in prediction accuracy. In D3, Fisher's exact test shows a significant difference in precision, while McNemar's test shows a significant difference in prediction accuracy; recall does not differ significantly. Overall, LEREDD and BERT show generally comparable performance, with significant differences in selected metrics in D2 and D3. However, given LEREDD's superior performance in detecting specific dependency types and the training costs associated with BERT's reliance on large labeled datasets, LEREDD offers practical advantages over BERT.

\paragraph{\textbf{Robustness Analysis}}
LEREDD demonstrated stable performance across repeated executions. The average standard deviation was only 0.020, despite the very limited number of instances for several dependency types. In fact, the standard deviation exceeded 0.05 in only five cases, all of which corresponded to minority dependency types (\textit{Requires}, \textit{Implements}, and \textit{Conflicts}). These results suggest that LEREDD produces consistent predictions even under severe class imbalance, further demonstrating its robustness.

\begin{formal}
\textbf{Answer to RQ4:} In the intra-dataset setup, LEREDD achieves higher accuracy and $F_1$-scores compared to the baselines for both \textit{No dependency} cases and specific dependency types. It yields relative gains of 33.33\% over TF-IDF \& LSA and 10.00\% over fine-tuned BERT in overall $F_1$-scores, increasing to 103.13\% and 25.00\%, respectively, for specific dependency types. LEREDD also provides more stable performance across systems. Its advantage over TF-IDF \& LSA is statistically significant, whereas the difference with BERT is not. Considering BERT's training costs and reliance on large datasets, LEREDD represents a more practical alternative.
\end{formal}

\subsection{\textbf{RQ5. Comparison of LEREDD with Baselines for Cross-Dataset Dependency Detection}}\label{sec:RQ5Results}

To address this RQ, we evaluate LEREDD's performance against baselines using a cross-dataset experimental setup: one dataset is used for testing, while a different dataset serves as the training set for BERT and as an example pool for LEREDD. Because the TF-IDF \& LSA baseline requires no training, we report the same experiments conducted for RQ3, with performance metrics computed over the entire requirements set. For fine-tuned BERT and LEREDD, we run six experiments each to cover all possible training–testing combinations across the three datasets, for a total of 12 experiments. For LEREDD, we experiment with two variants: with and without RAG context. The latter assesses whether incorporating domain context improves LEREDD's performance when retrieved examples come from a different dataset, as they may not provide relevant domain information. For LEREDD with RAG, we use the configuration identified in RQ2: 6 chunks of 1000 characters with a 200-character overlap.  

\autoref{tab:CrossDatasetResults} reports the experimental results for LEREDD and the baselines. To facilitate direct comparison in \autoref{tab:CrossDatasetResults}, we duplicate the TF-IDF \& LSA results for combinations sharing the same testing dataset. For each training–testing combination, we highlight the highest-performing results in bold. Note that the total number of instances and the dependency types covered by each combination vary, as they depend on: (1) whether the training set contains sufficient examples to support LEREDD (i.e., six examples per dependency type), and (2) whether the test set includes instances of the corresponding dependency types. For instance, the D2-D1 combination covers the \textit{Implements} dependency type, whereas D3-D1 does not. Although both combinations use D1 as the test set, D2 contains more than six training instances for the \textit{Implements} dependency type, while D3 does not (see \autoref{tab:dataset}).

\begin{table*}[t]
\centering
\scriptsize
\caption{Experimental Results of LEREDD Against Baselines for Cross-Dataset Evaluation across Three Datasets}
\label{tab:CrossDatasetResults}
\resizebox{\textwidth}{!}{
\begin{tabular}{c|c|c|c|ccc|c|ccc|c|ccc|c|ccc|c}
\hline
\multirow{2}{*}{\textbf{Training}} & \multirow{2}{*}{\textbf{Testing}} & \textbf{Dependency} & \multicolumn{4}{c|}{\textbf{TF-IDF \& LSA}} & \multicolumn{4}{c|}{\textbf{fine-tuned BERT}} & \multicolumn{4}{c|}{\textbf{LEREDD}} & \multicolumn{4}{c|}{\textbf{LEREDD with RAG}} & \multirow{2}{*}{\textbf{Supp}} \\ \cline{4-19} 
 &  & \textbf{Type} & \textbf{Acc} & \textbf{P} & \textbf{R} & \textbf{F1} & \textbf{Acc} & \textbf{P} & \textbf{R} & \textbf{F1} & \textbf{Acc} & \textbf{P} & \textbf{R} & \textbf{F1} & \textbf{Acc} & \textbf{P} & \textbf{R} & \textbf{F1} &  \\ \hline

 \multirow{4}{*}{D2} & \multirow{4}{*}{D1} &  No Dependency& \multirow{3}{*}{0.70 } & 0.86 & 0.76 & 0.81 & \multirow{3}{*}{0.67 } & 0.84 & 0.79 &0.82  & \multirow{3}{*}{0.84  } &0.86  &0.98 &\textbf{0.92} &  \multirow{3}{*}{\textbf{0.86} } &0.87  &0.98 &\textbf{0.92} &  229\\
 &  & Requires &  & 0.18 & 0.30 & 0.23 &  & 0.16 & 0.28 & 0.20 &  &0.68  &0.21 &0.32  & &0.83  &0.27  &\textbf{0.40}  &  40\\ 
  &  & Implements &  & N/A & N/A & N/A &  & 0.00 & 0.00 & 0.00 &  &0.64  &0.48 &0.55  & &0.72   &0.67   &\textbf{0.69 } &  18\\ \cline{3-3} \cline{5-7} \cline{9-11} \cline{13-15} \cline{17-20} 

 &  & \textit{Overall} &  &0.76  & 0.69 & 0.72 &  & 0.69 & 0.67 & 0.68 &  &0.82  &0.84   &0.81    & &\textbf{0.85}    &\textbf{0.86}    &\textbf{0.83}   & 287 \\ \hline
 
  \multirow{3}{*}{D3} & \multirow{3}{*}{D1} &  No Dependency& \multirow{3}{*}{ 0.70 } & 0.86 & 0.76 & 0.81 & \multirow{3}{*}{ 0.35} & 0.88 & 0.28 & 0.42 & \multirow{3}{*}{0.89  } &0.91   &0.97 &\textbf{0.94}  & \multirow{3}{*}{\textbf{0.90}  }&0.91   &0.98   &\textbf{0.94}   &  229\\
 &  &  Requires &  & 0.18 & 0.30 & 0.23 &  & 0.16 & 0.78 & 0.26 &  &0.70 &0.44  &0.54 & &0.77 &0.45 &\textbf{0.57}  & 40\\ 
   \cline{3-3} \cline{5-7} \cline{9-11} \cline{13-15} \cline{17-20}

 &  & \textit{Overall} &  &0.76  & 0.69 & 0.72 &  & 0.77 & 0.35 & 0.40 &  &0.88  &0.89  &\textbf{0.88}   & &\textbf{0.89}   &\textbf{0.90}    &\textbf{0.88}& 269 \\
 
 \hline
 
 \multirow{4}{*}{D1} & \multirow{4}{*}{D2} &  No Dependency& \multirow{4}{*}{ 0.78} & 0.96 & 0.79 & 0.86 & \multirow{4}{*}{0.64 } & 0.86 & 0.74 & 0.79 & \multirow{4}{*}{0.89} &0.90  &0.98 &0.94  &\multirow{4}{*}{\textbf{0.90}  } &0.93  &0.99  &\textbf{0.96}   &  170\\
 &  &  Requires &  & 0.23 & 0.65 & 0.34 &  & 0.00 & 0.00 & 0.00 &  &0.89  &0.27  &0.42 & &0.70 &0.41   &\textbf{0.52}  & 17 \\
   &  & Implements &  & N/A & N/A & N/A &  & 0.33 & 0.10 & 0.15 &  &0.62  &0.43  &0.50  & &0.94  &0.40 &\textbf{0.55}   &  10\\
    &  & Details &  & N/A & N/A & N/A &  & 0.00 & 0.00 & 0.00 &  &0.28 &0.33 &0.30  &  &0.26  &0.50  &\textbf{0.34}  & 2 \\\cline{3-3} \cline{5-7} \cline{9-11} \cline{13-15} \cline{17-20} 

 &  & \textit{Overall} &  & 0.89 & 0.78 & 0.81 &  & 0.75 &  0.64 & 0.69  &  & 0.88  & 0.88 &0.87   & &\textbf{0.90}   &\textbf{0.90}  &\textbf{0.89}   & 199 \\ \hline 

  \multirow{3}{*}{D3} & \multirow{3}{*}{D2} &  No Dependency& \multirow{3}{*}{0.78 } & 0.96 & 0.79 & 0.86 & \multirow{3}{*}{0.46 } & 0.95 & 0.43 & 0.59 & \multirow{3}{*}{\textbf{0.96} } &0.97  &0.98  &\textbf{0.98}  & \multirow{3}{*}{\textbf{0.96} }&0.97   &0.99   &\textbf{0.98}  & 170\\
 &  &  Requires &  & 0.23 & 0.65 & 0.34 &  & 0.12 & 0.76 & 0.20 &  &0.83  &0.69  &0.75  & &0.86  &0.71   &\textbf{0.77}   & 17 \\ \cline{3-3} \cline{5-7} \cline{9-11} \cline{13-15} \cline{17-20} 

 &  & \textit{Overall} &  & 0.89 & 0.78 & 0.81 &  & 0.87 & 0.46 & 0.56 &  &\textbf{0.96}   & \textbf{0.96}    &\textbf{0.96}  & &\textbf{0.96}   &\textbf{0.96}    &\textbf{0.96}   & 187 \\ \hline
 
 \multirow{4}{*}{D1} & \multirow{4}{*}{D3} &  No Dependency& \multirow{4}{*}{ 0.53} & 0.85 & 0.47 & 0.60 & \multirow{4}{*}{ 0.72} & 0.73 &  0.98 & 0.84 & \multirow{4}{*}{0.74 } &0.80  &0.94   &0.86  &\multirow{4}{*}{\textbf{0.75}} &0.88   &0.92  &\textbf{0.90}   &  144\\
 &  &  Requires &  & 0.30 & 0.73 & 0.43 &  & 0.00 & 0.00 & 0.00 &  &0.58  &0.23  &0.33  & &0.66   &0.34  &\textbf{0.45}  &  45\\
  &  &  Implements &  & N/A & N/A & N/A &  & 0.00 & 0.00 & 0.00 &  &0.00  & 0.00 & 0.00   & &0.00   &0.00  &0.00  & 3 \\
  &  & Details &  & N/A & N/A & N/A &  & 0.00 & 0.00 & 0.00 &  &  0.00  & 0.00  & 0.00  & &0.00   &0.00   &0.00  & 1 \\
 &  &  Conflicts &  & N/A & N/A & N/A &  & 0.00 &  0.00 & 0.00 &  &1.00  &0.33  &\textbf{0.49}  & &0.00   &0.00   &0.00  & 4 \\ \cline{3-3} \cline{5-7} \cline{9-11} \cline{13-15} \cline{17-20}  

 &  & \textit{Overall} &  & 0.72 & 0.53 & 0.56 &  & 0.53 & 0.72 & 0.61 &  &0.74 &0.74  &0.74  & &\textbf{0.79}   &\textbf{0.75}   &\textbf{0.76}  & 197 \\ \hline

\multirow{3}{*}{D2} & \multirow{3}{*}{D3} &  No Dependency& \multirow{3}{*}{ 0.53} & 0.85 & 0.47 & 0.60 & \multirow{3}{*}{0.49 } & 0.85 & 0.40 & 0.55 & \multirow{3}{*}{0.78 } &0.84  &0.92  &0.88  &\multirow{4}{*}{\textbf{0.81} } &0.90  &0.94  &\textbf{0.92}  &  144\\
 &  &  Requires &  & 0.30 & 0.73 & 0.43 &  & 0.29 & 0.80 & 0.43 &  &0.62 &0.39  &0.47  & &0.80  &0.45  &\textbf{0.58} & 45 \\ 
   &  &  Implements &  & N/A & N/A & N/A &  & 0.00 & 0.00 & 0.00 &  &0.18 &0.33  &\textbf{0.23} & &0.00  &0.00 &0.00 &  3 \\
\cline{3-3} \cline{5-7} \cline{9-11} \cline{13-15} \cline{17-20}  

 &  & \textit{Overall} &  & 0.72 & 0.53 & 0.56 &  & 0.71 & 0.49 & 0.51 &  &0.77  &0.79  &0.77  & &\textbf{0.86}  &\textbf{0.81}  &\textbf{0.82}  & 192 \\ \hline
 
\end{tabular}}
\footnotesize

Note: N/A indicates dependency types not supported by TF-IDF \& LSA.
\end{table*}

\paragraph{\textbf{Overall Performance}} Both LEREDD variants outperformed the baselines in overall accuracy, precision, recall, and $F_1$-score, with LEREDD with RAG achieving the best overall performance across all dataset combinations. Specifically, it achieved overall $F_1$-scores ranging from 0.76 for the D1-D3 combination to 0.96 for the D3-D2 combination.
While LEREDD with RAG achieved the highest overall $F_1$-score, averaging 0.86 across all combinations, LEREDD without RAG followed closely, averaging 0.84. In contrast, the baselines achieved substantially lower $F_1$-scores, with TF-IDF \& LSA and BERT averaging 0.69 and 0.57, respectively. These results highlight how sensitive ML-based approaches are to the characteristics of the training dataset. 

Compared to the highest-performing baseline, TF-IDF \& LSA, which achieved the best performance in five out of the six combinations, LEREDD achieved relative $F_1$-score improvements of 15.28\%, 22.22\%, 9.88\%, 18.52\%, 33.93\%, and 46.43\% across the six combinations. Although TF-IDF \& LSA identified only the \textit{Requires} dependency type, the most frequent dependency type across datasets, LEREDD consistently outperformed it. This further demonstrates LEREDD's advantage in handling diverse dependency types. Furthermore, compared to RQ4, the performance decrease in RQ5 remained limited, with both the average accuracy and $F_1$-score decreasing by only 2.27\%.

\paragraph{\textbf{Performance on Specific Dependency Types}} 

As before, both LEREDD variants achieved very high $F_1$-scores for \textit{No dependency}. The two variants achieved identical results in three of the six dataset combinations, with $F_1$-scores of 0.92, 0.94, and 0.98. For the remaining three combinations, LEREDD with RAG slightly outperformed LEREDD without RAG, improving the $F_1$-score from 0.94 to 0.96 for D1-D2, from 0.86 to 0.90 for D1-D3, and from 0.88 to 0.92 for D2-D3. These results demonstrate LEREDD's consistent performance in identifying non-dependent requirement pairs. 
The two baselines also achieved relatively high $F_1$-scores for the \textit{No dependency} class in most cases. However, their performance was less consistent than LEREDD's. For TF-IDF \& LSA, the results were strong when D1 and D2 were used as testing sets, achieving $F_1$-scores of 0.81 and 0.86, respectively, but decreased to 0.60 when D3 was used for testing. Similarly, fine-tuned BERT showed substantial fluctuations, even when evaluated on the same test set. For example, for D1, the $F_1$-score for \textit{No dependency} decreased from 0.82 to 0.42 when D2 and D3 were used as training datasets, respectively. Similarly, it decreased from 0.79 to 0.59 for D2 when D1 and D3 were used for training, and from 0.84 to 0.55 for D3 when D1 and D2 were used for training. These variations contributed to the differences in overall $F_1$-scores across combinations using the same testing dataset.

For specific dependency types, although LEREDD achieved lower $F_1$-scores than the intra-dataset evaluation results, it still substantially outperformed the baselines. LEREDD with RAG outperformed LEREDD without RAG for almost all dependency types. In fact, RAG's main advantage is improved detection of specific dependency types, as the $F_1$-scores for \textit{No dependency} were either identical or relatively close between the two variants. For instance, in the D2-D1 combination, incorporating RAG improved the $F_1$-score for \textit{Requires} from 0.32 to 0.40 and for \textit{Implements} from 0.55 to 0.69. LEREDD without RAG outperformed LEREDD with RAG in only two instances, both corresponding to dependency types with extremely low support: \textit{Conflicts} in D1-D3 (4 instances) and \textit{Implements} in D2-D3 (3 instances). These results suggest that while RAG also improves \textit{No dependency} detection, its primary contribution lies in improving the identification of specific dependency types by providing relevant domain context.

In contrast, both baselines achieved substantially lower performance on specific dependency types than in previous research questions. TF-IDF \& LSA and fine-tuned BERT achieved average $F_1$-scores of 0.33 and 0.18 for \textit{Requires} across experiments, respectively, compared to 0.55 for LEREDD with RAG. For other dependency types, BERT frequently achieved an $F_1$-score of 0.00, indicating its difficulty in identifying specific, less frequent dependency categories. Although LEREDD also obtained an $F_1$-score of 0.00 in some cases, this occurred less frequently than for BERT (4 instances compared to 8) and was limited to dependency types with very low support (4 instances or fewer).
This extreme decline is attributable to fine-tuned BERT, like most ML models, being highly dependent on similarities between training and test data. Because these models learn patterns and assign class weights from the training corpus, using training and testing sets from different distributions substantially affects their performance. These results highlight the sensitivity of ML-based approaches to training-data characteristics, making them less suitable for dependency detection in real-world settings with sparse and inter-system dependency annotations.

Overall, using RAG with LEREDD achieves higher accuracy and $F_1$-scores in the inter-dataset setting, significantly outperforming the baselines in cross-dataset evaluation. These results are particularly promising, given the scarcity of training data and lack of annotated requirements from the same dataset in real-world settings.

\paragraph{\textbf{Confidence Score Analysis}}

Given the slight decrease in $F_1$-scores observed between RQ4 and RQ5, it is important to analyze the confidence scores from intra- and inter-dataset evaluations. In the intra-dataset evaluation, a confidence score of 5 was assigned to 95.83\% of predictions for D1, 89.17\% for D2, and 90.60\% for D3. In comparison, LEREDD with RAG achieved a confidence score of 5 for only 66.08\% of predictions for D1, 69.05\% for D2, and 54.31\% for D3 in the inter-dataset evaluation. These results show a substantial decrease in LEREDD's confidence across all datasets when transitioning from intra-dataset to inter-dataset evaluation. This suggests that LEREDD is more confident when given examples from the same domain. Although incorporating domain context via RAG improved inter-dataset performance, it did not substantially increase LEREDD's confidence scores, indicating that retrieved domain information primarily improved prediction accuracy and $F_1$-scores rather than model certainty.

\paragraph{\textbf{Statistical Testing}}

When comparing LEREDD with LEREDD with RAG, we observed no statistically significant differences across dataset combinations, as all Fisher's exact and McNemar's test p-values exceeded 0.05. Nevertheless, performance differed by dependency type, favoring LEREDD with RAG. Therefore, the following analysis compares LEREDD with RAG against the baselines to determine whether it provides a statistically significant advantage.

Across the inter-dataset evaluations, LEREDD consistently outperforms TF-IDF \& LSA, with statistically significant differences in precision, recall, and prediction accuracy across all dataset combinations. Overall, these results provide strong and consistent evidence that LEREDD significantly outperforms TF-IDF \& LSA in inter-dataset dependency detection.

Compared with BERT, LEREDD generally demonstrates significantly superior accuracy. 
Overall, LEREDD consistently outperforms BERT across inter-dataset evaluations, with statistically significant improvements in precision and paired prediction accuracy across all six dataset combinations, and in recall across five of the six combinations.

\paragraph{\textbf{Robustness Analysis}}

Both variants of LEREDD demonstrated high robustness across experiments. LEREDD without RAG achieved an average standard deviation of 0.024, while LEREDD with RAG achieved an average standard deviation of 0.013. Although both variants were highly robust, introducing RAG reduced the average deviation by 46\%. In fact, LEREDD with RAG exhibited deviations above 0.05 in only five instances, with the highest deviation reaching 0.08, compared to eight instances for LEREDD, with a highest deviation of 0.24. These high deviations were exclusively related to specific dependency types with very low support (e.g., \textit{details} for D1-D2 with only 2 instances), where a single different prediction can substantially affect the deviation. These results suggest that introducing domain context not only improved LEREDD's confidence but also enhanced its robustness, likely because the retrieved context gave the LLM more relevant domain information and enabled more informed, grounded predictions.

\begin{formal}
\textbf{Answer to RQ5:} In the inter-dataset setup, LEREDD with RAG outperforms LEREDD with few-shot alone. LEREDD with RAG statistically significantly outperforms the baselines in terms of precision, recall, and overall accuracy, both for \textit{No dependency} and for specific dependency types. It achieves relative gains in overall $F_1$-scores of 22.86\% and 50.88\% over TF-IDF \& LSA and BERT, respectively, increasing to 280\% over BERT for specific dependency types. Furthermore, its performance remains comparable to the intra-dataset setup, with average accuracy and $F_1$ decreasing by only 2.27\%.
\end{formal}

\subsection{\textbf{RQ6. Industrial Experts' Agreement with LEREDD Predictions}}\label{sec:RQ6Results}

To evaluate the level of agreement between industrial experts and LEREDD's predictions, we provided the experts with LEREDD's outputs and asked them to review the predictions and correct them if they considered them incorrect. \autoref{tab:IndustrialDataset} presents the results of LEREDD evaluated against the ground truth, which consists of the experts' final annotations.

\begin{table}[t]
\centering
\tiny
\caption{Performance of LEREDD on the industrial dataset}
\label{tab:IndustrialDataset}
\resizebox{\columnwidth}{!}{
\begin{tabular}{c|c|ccc|c}
\hline

\textbf{Dependency} & \multicolumn{4}{c|}{\textbf{LEREDD}} & \multirow{2}{*}{\textbf{Supp}} \\ \cline{2-5}
\textbf{Type} & \textbf{Acc} & \textbf{P} & \textbf{R} & \textbf{F1} &   \\ \hline

No Dependency  & \multirow{6}{*}{0.92} & 1.00 & 0.88 & 0.93 & 40\\
Requires       &                  & 0.90 & 1.00 & 0.95 & 26\\
Implements     &                  & 0.90 & 0.93 & 0.92 & 30\\
Details        &                  & 0.94 & 0.89 & 0.92 & 19\\
Conflicts      &                  & 0.33 & 1.00 & 0.50 & 1\\
\cline{1-1} \cline{3-6}
\textit{Overall} &                & 0.94 & 0.92 & 0.93 & 116 \\
\hline

\end{tabular}}
\end{table}

As the table shows, the experts' final annotations largely agree with LEREDD's predictions, with average accuracy, precision, recall, and $F_1$-score reaching 0.92, 0.94, 0.92, and 0.93, respectively. In fact, scores for almost all metrics and dependency types were at or above 0.9, except for the Conflicts dependency, which may be explained by the presence of only one instance of this dependency type in the dataset. These results indicate strong agreement between the experts' final annotations and LEREDD's predictions, demonstrating LEREDD's ability to distinguish dependent from non-dependent requirements and identify specific dependency types in an industrial setting.

Beyond the predictions themselves, the experts also found LEREDD's rationales and confidence scores useful during the review process. Specifically, all experts noted that the rationales greatly facilitated examining LEREDD's predictions by providing deeper insights into the reasoning behind each prediction. One expert also described the reasoning presented in the rationales as interesting and well-founded. Furthermore, the confidence scores provided additional guidance during the review. As one expert noted, lower confidence scores indicated predictions for which LEREDD was less certain, drawing attention to requirement pairs that were more likely to be ambiguous or misclassified.

\subsubsection{Analysis of Unanimous and Majority Agreement}

When comparing the results for requirement pairs with unanimous expert agreement and those for which majority voting was applied, a pattern emerges. Experts unanimously agreed on most \textit{No dependency} annotations (29 out of 40), whereas most specific dependency types were identified through majority agreement. Specifically, 17 of 30 Implements, 16 of 26 Requires, 11 of 19 Details, and the only Conflicts instance were identified through majority agreement. These results suggest that distinguishing between dependent and non-dependent requirement pairs is less challenging for experts than identifying specific dependency types.

Multiple dependency types can co-exist within the same requirement pair. However, for cost-efficiency reasons, our annotation process allowed only one dependency type, the most evident one, to be assigned to each pair. This constraint may explain the disagreements among experts regarding specific dependency types. An examination of the majority-voting results revealed that \textit{Implements} was the dependency type most frequently involved in disagreements, with 11 cases involving \textit{Details} and six involving \textit{Requires}. One possible explanation is that these dependency types are likely to co-exist within the same requirement pair, leading experts to identify different types in such cases. For example, a lower-level requirement can both implement a higher-level requirement and provide additional details about the functionality it describes. Similarly, a lower-level requirement can both implement a higher-level requirement and specify a functionality that is a prerequisite for its fulfillment. These findings further suggest that, even among experts, determining specific dependency types is more challenging than determining whether a dependency exists.

Beyond the annotations themselves, we further examined the confidence scores assigned by LEREDD in relation to unanimous and majority agreement. For requirement pairs with unanimous agreement, LEREDD assigned a confidence score of 5 in 48\% of cases, 4 in 48\%, and 3 in 4\%. In contrast, for pairs with disagreements, it assigned a score of 5 in 12\% of cases, 4 in 72\%, and 3 in 16\%. These results indicate that higher confidence scores were more frequently associated with unanimous expert agreement, as reflected by a higher proportion of score 5 and a lower proportion of score 3. Conversely, lower confidence scores were more common among pairs with disagreements, as reflected by a lower proportion of score 5 and a higher proportion of score 3. This observation further supports our previous finding that LEREDD-generated confidence scores can help guide experts during manual verification.

Overall, manual verification remains necessary when applying LEREDD. The high agreement between the experts' final annotations and LEREDD's predictions demonstrates the reliability of its results. Together with the reported usefulness of its rationales and confidence scores, these findings support the applicability of LEREDD in industrial settings, including safety-critical contexts, where human review remains an essential part of the process.

\begin{formal}
\textbf{Answer to RQ6:} Industrial experts show high agreement with LEREDD’s predictions, with LEREDD achieving an average $F_1$-score of 92\% against the experts’ final annotations. LEREDD performed well both in detecting the No dependencies and in classifying specific dependency types, with $F_1$-scores generally above 0.90. Moreover, the rationales and confidence scores LEREDD generated helped support the experts’ manual verification.
\end{formal}

\section{Discussion}\label{sec:discussion}

In this paper, we introduce LEREDD, an LLM-based approach for automated detection of requirement dependencies. Through extensive evaluation across multiple datasets and experimental settings, we demonstrated that LEREDD consistently outperforms SOTA baselines while maintaining strong generalization. 

\paragraph{\textbf{\textit{Zero-Shot Limitations}}}
Zero-shot GPT-5.4 reliably detects \textit{No dependency} but performs poorly on specific dependency types  (e.g., \textit{Requires}, \textit{Details}). This indicates that dependency detection is not merely semantic-similarity matching but a structured form of reasoning that requires domain context.
Without task-specific guidance, zero-shot models rely on coarse semantic signals, tend to over-predict dominant classes, and collapse on minority dependencies. 
These findings indicate that zero-shot LLMs are insufficient for detecting specific dependency types.

\paragraph{\textbf{\textit{Few-Shot Prompting: Retrieval Quality over Quantity}}}
Our evaluation indicates that few-shot prompting is the most effective strategy for requirement dependency detection, particularly when the retrieved examples originate from the same system, for instance, when new requirements are added to an already annotated set of requirements. This is because such examples provide both task-specific guidance and relevant contextual information. 
The experiments show that providing six examples per dependency type is sufficient to guide the model through the task, suggesting that retrieval quality is more important than retrieval scale.

\paragraph{\textbf{\textit{RAG: Structured Context Augmentation}}} 
In intra-dataset settings, adding RAG to few-shot prompting does not provide a significant advantage and can reduce performance in some cases. This suggests that the retrieved examples already provide sufficient context, and that additional RAG may introduce unnecessary noise. In contrast, in inter-dataset settings, RAG provides an advantage over Few-shot alone, particularly for specific dependency types. This is because examples from different systems, while providing task-specific guidance, lack the domain-specific information needed to identify dependencies. With RAG, the optimal performance at moderate chunk sizes highlights the trade-off between contextual coverage and noise. Dependency detection therefore benefits from structured and selective context augmentation rather than unrestricted information injection, particularly in inter-dataset settings.

\paragraph{\textbf{\textit{LEREDD: Robustness Within and Across Datasets}}}
LEREDD maintains strong, stable performance in both intra- and cross-dataset evaluations. It achieves superior results on minority dependencies and on \textit{No dependency} cases. This advantage persists under a distribution shift, where fine-tuned BERT degrades substantially.
LEREDD leverages dynamic retrieval and inferential reasoning, reducing dependence on static training distributions. Consequently, it maintains higher, more balanced precision and recall across systems, demonstrating stronger generalization in evolving industrial contexts. The experiments further show that LEREDD produces consistent results across runs, with only small deviations, indicating stable performance despite the inherent stochasticity of LLMs.
Regarding specific dependency types, while LEREDD offers substantial improvements over baselines and zero-shot LLMs, further refinement is necessary to achieve optimal performance. LEREDD is particularly proficient in identifying \textit{No dependency} cases. Because these cases constitute the vast majority of instances, filtering them out substantially reduces the overhead of dependency analysis and is of high practical value. 

\paragraph{\textbf{\textit{LEREDD: Computational Cost}}}
In terms of computational time, LEREDD offers a good trade-off between BERT and TF-IDF \& LSA while achieving much higher and more stable accuracy. Our empirical results show that TF-IDF \& LSA required an average of 2.48 seconds, while fine-tuned BERT and LEREDD required an average of 4 minutes 3 seconds and 1 minute 48 seconds, respectively, across the three systems during the intra-dataset evaluation.  

\paragraph{\textbf{\textit{LEREDD: Applicability in Industrial Settings}}} The high level of agreement between industrial partners and LEREDD's predictions indicates that LEREDD captures the relevant domain knowledge and expertise required for this task, both in identifying the existence of dependencies and in classifying their types in industrial requirements. This suggests that LEREDD can identify meaningful relationships between requirements. Its generated rationales provide transparency into the basis of its predictions, while confidence scores help analysts assess prediction reliability and prioritize manual verification. These are particularly valuable in safety-critical systems, where expert verification remains essential. 
Overall, our results suggest that LEREDD has the potential to support dependency detection in industrial settings by partially automating the process and reducing the effort required to examine large sets of requirements. However, since our industrial evaluation was conducted on a single dataset, further experiments on additional industrial datasets are needed to confirm this finding.

\paragraph{\textbf{\textit{Implications}}} Our findings suggest several important insights into using LLMs for requirement dependency detection. First, the accuracy of detecting \textit{No dependency} cases is so high that it can be extremely useful for filtering out most requirement pairs when analyzing inter-requirement dependencies. Second, zero-shot LLMs are insufficient for accurately identifying specific dependency types and often yield low $F_1$-scores even in \textit{No dependency} cases, underscoring the need for task-specific guidance. Third, a relatively large set of carefully selected, relevant examples can yield strong results. Fourth, for RAG, retrieval precision matters more than retrieval volume, as highly relevant context yields notable gains in $F_1$-scores beyond those achieved with few-shot prompting alone. Finally, LLMs' inferential capabilities, combined with their reduced reliance on dataset-specific distributions, improve robustness across systems and enable more consistent performance in cross-dataset evaluation settings.

From an industrial perspective, this study provides evidence of LLMs' potential to support the complex, recurring task of detecting requirement dependencies. By semi-automating dependency detection, LEREDD can reduce the effort required of requirements engineers and facilitate systematic consideration of dependencies that might otherwise be overlooked. Engineers can focus on the smaller set of requirements LEREDD identifies as related, reducing manual analysis workload. 
Given the evolving nature of requirements in most practical settings, LEREDD provides a maintainable and efficient approach to integrating dependency detection more systematically into the software development process. Finally, as an LLM-based approach, LEREDD can be integrated into existing agent-based architectures, facilitating adoption in industrial development workflows.

\section{Threats to validity}\label{sec:threatsToValidity}

\textit{\textbf{Construct Validity.}}
One threat to construct validity arises from the annotation of the public datasets, which may be biased and error-prone, particularly for specific dependencies. To mitigate this risk, we followed predefined annotation guidelines and had two independent annotators cross-validate all annotations to reduce potential bias.  Furthermore, selecting pairs for manual annotation based on semantic similarity yields a higher proportion of dependencies than random selection under the same labeling budget. However, this was intentional, as semantic similarity was used solely to reduce annotation effort while ensuring a higher number and greater diversity of dependencies in the datasets. It did not influence the annotation process or LEREDD’s predictions. Replicating baselines may also raise validity concerns if implementations differ. To reduce this risk, we reused the parameters from the original approaches and tuned hyperparameters to identify optimal configurations, ensuring fair, consistent comparisons.

\textit{\textbf{Internal Validity.}}
A relevant threat to internal validity concerns the design and parameter choices made for LEREDD, including prompting techniques, the number of in-context examples, and the selected models. To mitigate this risk, we determined these configurations through the extensive empirical evaluations detailed in Section~\ref{sec:Results}. Furthermore, LLM-based methods may produce varying outputs due to their stochastic nature. To ensure reproducibility and minimize randomness as a confounding factor, we fixed the temperature at 0.2 and repeated each experiment three times to improve the robustness and reliability of the results. We also analyzed the variation across runs, and the standard deviation was generally negligible when applying LEREDD. Finally, to mitigate the threat that observed performance differences between LEREDD and the baseline approaches are attributable to random variation, we performed statistical significance tests, including Fisher's exact test and McNemar's test. 

\textit{\textbf{External Validity.}}
Regarding external validity, our datasets are confined to the automotive domain, which may limit generalizability to other application areas. However, requirements engineering practices are similar across many regulated domains that must comply with functional safety standards. Indeed, the automotive ISO 26262 standard is a specialization of IEC 61508, which provides a generic foundation for functional safety standards, including requirements engineering. Furthermore, the generalizability of our findings may be limited to the evaluation datasets. To alleviate these concerns, we selected three systems that differ in size and distribution, included an industrial dataset in the evaluation, and further conducted cross-dataset evaluations. Nevertheless, further experiments involving systems from different domains are needed to better assess the generalizability of our findings.  


Another potential threat concerns the LLM used in our approach and evaluation. We use GPT-5.4, selected for its demonstrated effectiveness and widespread use in requirements engineering. Using other LLMs may produce different results. Nevertheless, we expect the overall conclusions to remain similar across models with comparable capabilities, as our multi-model comparison suggests. As future work, we plan to evaluate additional proprietary LLMs to further assess the generalizability of our findings.

\textit{\textbf{Reliability.}}
To mitigate reliability threats and ensure reproducibility, we provide a replication package that includes the source code, configurations, and datasets used throughout the paper. 
The replication package will be made available upon publication 

\section{Conclusion}\label{sec:Conclusion}
In this paper, we introduce LEREDD, an LLM-based approach for automatically detecting various types of dependencies between NL requirements. Detecting such dependencies is essential for supporting activities such as requirements change management, development, and testing. LEREDD leverages RAG and ICL to retrieve domain-specific contextual information that systematically guides the LLM during detection. We empirically evaluate LEREDD on four datasets comprising 929 requirement pairs and compare it to baselines under both intra-system and inter-system settings for few-shot example selection.

The experimental results show that LEREDD significantly outperforms existing baselines, particularly for specific dependency types. It achieves an average $F_1$-score of 88\% across the three systems and all dependency types. The performance is particularly strong for the \textit{No dependency} class, which reaches an average $F_1$-score of 93\%. These results are especially compelling because \textit{No dependency} cases account for the vast majority of requirement pairs. By efficiently filtering these cases, LEREDD can substantially reduce the effort and time required for dependency analysis. Notably, LEREDD demonstrates robustness in inter-system evaluations, achieving a significantly higher $F_1$-score than the baselines, while its average $F_1$-score decreases by only 2.27\% compared with the intra-system setting. Furthermore, an evaluation of LEREDD on an industrial dataset revealed a high level of similarity between its predictions and those of industrial experts, demonstrating the reliability of its predictions on industrial requirements. The evaluation also showed that the generated rationales and confidence scores help facilitate the analysis and verification of LEREDD's results. Finally, the annotated corpus provided in this paper can serve as a valuable resource for training and benchmarking for future research. 

As future work, we will evaluate LEREDD on additional industrial datasets to further assess its generalizability. We will also extend LEREDD  to identify multiple dependencies within a requirement pair, as well as indirect and implicit dependencies among requirements. Finally, we will investigate how predicted dependencies can support impact analysis when requirements evolve over time.

\section{Acknowledgments}

This work was supported by a research grant from General Motors, as well as the Canada Research Chair and Discovery Grant programs of the Natural Sciences and Engineering Research Council of Canada (NSERC). Lionel C. Briand’s contribution was partially funded by the Research Ireland grant 13/RC/209.

\bibliographystyle{IEEEtran}
\bibliography{biblio}

\end{document}